\documentclass{aa}  

\pdfoutput=1
\usepackage{natbib,twoopt}
\usepackage[breaklinks=true]{hyperref} 
\bibpunct{(}{)}{;}{a}{}{,} 
\makeatletter
\newcommandtwoopt{\citeads}[3][][]{\href{http://adsabs.harvard.edu/abs/#3}%
{\def\hyper@linkstart##1##2{}%
\let\hyper@linkend\@empty\citealp[#1][#2]{#3}}}
\newcommandtwoopt{\citepads}[3][][]{\href{http://adsabs.harvard.edu/abs/#3}%
{\def\hyper@linkstart##1##2{}%
\let\hyper@linkend\@empty\citep[#1][#2]{#3}}}
\newcommandtwoopt{\citetads}[3][][]{\href{http://adsabs.harvard.edu/abs/#3}%
{\def\hyper@linkstart##1##2{}%
\let\hyper@linkend\@empty\citet[#1][#2]{#3}}}
\newcommandtwoopt{\citeyearads}[3][][]%
{\href{http://adsabs.harvard.edu/abs/#3}
{\def\hyper@linkstart##1##2{}%
\let\hyper@linkend\@empty\citeyear[#1][#2]{#3}}}
\makeatother
\usepackage{graphicx, subfigure}
\usepackage{tabularx}
\usepackage[varg]{txfonts}
\usepackage{rotating}
\usepackage{lscape}
\usepackage{pdflscape}
\def\xlinkspace#1 #2{%
 \ifx\relax#2%
 \xlinkdash#1-\relax
 \else
 \xlinkdash#1 -\relax
 \expandafter\xlinkspace\expandafter#2%
 \fi}

\def\xlinkdash#1-#2{%
 \ifx\relax#2%
 \tmp{#1}%
 \else
 \tmp{#1-}%
 \expandafter\xlinkdash\expandafter#2%
 \fi}

\makeatother

\begin{document} 

   \title{PyNeb: a new tool for analyzing emission lines}

   \subtitle{I. Code description and validation of results}

   \author{V. Luridiana
          \inst{\ref{iac} \and \ref{ull}}
          \thanks{e-mail: vale@iac.es}
          \and C. Morisset
          \inst{\ref{iaunam}}
          \and R. A. Shaw
          \inst{\ref{noao}}
          }
          \institute{Instituto de Astrof{\'\i}sica de Canarias, c/ V{{\'\i}a L\'actea s/n, E-38205 La Laguna, Tenerife, Spain}\label{iac}  
          \and{Departamento de Astrof{\'\i}sica, Universidad de La Laguna, E-38206, La Laguna, Tenerife, Spain}\label{ull}
          \and{Instituto de Astronom{\'\i}a, Universidad Nacional Aut\'onoma de M\'exico, Apdo. Postal 70264, 04510 M\'exico D.F., Mexico}\label{iaunam}
          \and{National Optical Astronomy Observatory, Tucson, AZ 85719, USA}\label{noao}}

   \date{}

  \abstract{
  Analysis of emission lines in gaseous nebulae yields direct measures 
   of physical conditions and chemical abundances and is the cornerstone of nebular astrophysics. 
   Although the physical problem is conceptually simple, its practical complexity can be overwhelming 
   since the amount of data to be analyzed steadily increases; furthermore, results depend crucially on the input atomic data, whose determination also improves each year. 
   To address these challenges we created PyNeb, an innovative code for analyzing emission lines. 
   PyNeb computes physical conditions and ionic and elemental abundances and produces both theoretical and observational diagnostic plots. 
   It is designed to be portable, modular, and largely customizable in aspects such as the atomic data used, 
   the format of the observational data to be analyzed, and the graphical output. 
   It gives full access to the intermediate quantities of the calculation, making it possible to write scripts tailored 
   to the specific type of analysis one wants to carry out. In the case of collisionally excited lines, 
   PyNeb works by solving the equilibrium equations for an \textit{n}-level atom; 
   in the case of recombination lines, it works by interpolation in emissivity tables. 
   The code offers a choice of extinction laws and ionization correction factors, which can be complemented by user-provided recipes. 
   It is entirely written in the python programming language and uses standard python libraries. 
   It is fully vectorized, making it apt for analyzing huge amounts of data.
   The code is stable and has been benchmarked against IRAF/NEBULAR. It is public, fully documented, and has already been satisfactorily used in a number of published papers.}


   \maketitle
%

\section{Introduction}

    The intensity of emission lines in the spectra of ionized nebulae carries valuable information on the physical conditions and chemical abundances of these objects. The physics of line formation in the relevant regime has been described in several seminal papers published throughout the past century 
\citep{1937ApJ....85..330M,1938ApJ....88...52B,1938ApJ....88..422B,1958MNRAS.118..477B,1989agna.book.....O} 
   and can be summarized in a handful of equations. They are conceptually simple, but cannot be solved analytically, so a numerical code is necessary to do the task. Furthermore, they depend on a number of parameters --- the atomic data --- which are only approximately known and whose determination improves steadily. 
   A suitable tool to carry out this task must therefore be prepared to include such advances. 
   
   This paper and a forthcoming one \citep[][hereinafter Paper II]{paperII} describe PyNeb, a new python package for the analysis of emission line spectra in ionized nebulae.
PyNeb (\url{http://www.iac.es/proyecto/PyNeb}) is the latest 
in a lineage of tools dedicated to addressing this problem. The first of these codes, FIVEL \citep{1987JRASC..81..195D}, is a FORTRAN program that solves the equilibrium equations for a five-level atom model to determine level populations and line emissivities, which are then used to apply simple temperature and density diagnostics. The heir of FIVEL, NEBULAR 
\citep{1995PASP..107..896S,1998ASPC..145..192S}, 
is an IRAF/STSDAS package that extends FIVEL's basic functionality with a small suite of 
tasks, such as determining ionic abundances and plotting diagnostic diagrams. NEBULAR also includes a simple 
tool to compute abundances from observed emission line intensities for over three dozen supported ions in a three-zone nebula ({low,} {medium,} and high-ionization) and a dereddening tool with a choice of interstellar extinction laws. One major asset of NEBULAR over FIVEL is that atomic data are no longer hardwired into the code; instead, they are read at run time from external files, which can therefore be conveniently updated without recompiling the source code. 
Both atomic data and observed emission line intensities 
are stored in FITS format to provide a standardized container for structured data with associated metadata. 
   Another asset of NEBULAR over FIVEL is that, for a given ion, users may employ any recognized line ratio by creating an expression for the desired transitions. Finally, a simple web interface was created for two of the simpler tasks, and it has proven to be one of its most popular uses. 
   
In its almost twenty years of life, NEBULAR has enjoyed 
wide use in the research community for analyzing physical conditions and chemical abundances in a variety of astrophysical contexts, including active galaxies, \ion{H}{ii} regions, and planetary nebulae (PNe). It has also been widely used as a starting point by people who model gaseous nebulae with photoionization codes.
The decision to recast NEBULAR in a modern and very popular programming 
language, python, was motivated by the desire to expose the internal functionality and data model to users through application interfaces (APIs), to provide a richer set of visualization and analysis tools, to provide 
easier implementation options for GUI and web-based user interfaces, and 
to make it simple for developers or advanced users to extend the original functionality of the code. The result of this effort is PyNeb, a code that, in addition to the functionality of NEBULAR, provides several new analysis features, a vast range of visualization tools, and full access to the intermediate quantities of the calculation. The latter feature makes it possible to write scripts tailored to the specific type of analysis one wants to carry out. 

In this paper, we give an overview of the current version of PyNeb \citep[v. 1.0.1; an earlier version of the code was described in ][]{2012IAUS..283..422L}. Specifically, Sect.~\ref{sec:physics} briefly reviews the physics underlying PyNeb; Sect.~\ref{sec:overview} provides an overview of the package functionality; 
Sect.~\ref{sec:code_validation} summarizes the validation of the code, consisting of a thorough comparison with NEBULAR; 
Sect.~\ref{sec:using} addresses practical aspects of the code, such as software requirements and documentation; 
finally, Sect.~\ref{sec:conclusions} recapitulates the present status of PyNeb and describes possible directions for expanding the code in the future. Two sample scripts illustrating how PyNeb can be used to solve real physical problems are included in the Appendix. Paper II will be entirely devoted to describing the new atomic database.

\section{Underlying physics}\label{sec:physics}

In photoionized objects in low-density environments, atoms are far away from LTE, and the only level that is appreciably populated is the ground state (with the only exception of excited metastable levels, such as the 2$^3$S of \ion{He}{i}, or excited levels of extremely low energy). The spectra emitted in nebular conditions are dominated by permitted lines of H and He ions and forbidden lines of ions of abundant metals, such as N, O, and Ne. Although in principle all line-formation processes are possible for all chemical species, in practice only some of them are effective for each given ion; thus, H and He lines mainly form in the downward cascade following recombination of H$^+$, He$^+$, and He$^{++}$, while the strongest metal lines mainly form in the radiative de-excitation following collisional excitation from the ground state. In both cases, the emissivity of a line is proportional to the product of the population of the upper level $n_u$ and the line transition probability $A_{ul}$. 
In the following, we summarize the main features of both types of lines.

\subsection{Hydrogen and helium: recombination lines} 

The largest contribution to H and He lines in ionized nebulae comes from recombination of  H$^+$, He$^+$, and He$^{++}$. The recombination cascade begins when a free electron recombines to an excited level and 
starts to fall through the atom's excited levels. The downward path is determined by the branching ratios from each level to the lower lying ones, possibly modulated by horizontal redistribution among sublevels of the same main level ({\it l}-mixing).
\cite{1987MNRAS.224..801H} and \cite{1995MNRAS.272...41S} performed detailed computations of the recombination cascade of H-like ions and made them available in tabular form for a wide array of temperature and density values. 

Although recombination is the most effective line-formation process for the ions of these two elements, the excited levels of H and He can also be populated through upward collisional excitations and continuum pumping. As anticipated above, both processes are mostly negligible in nebular conditions: the former is negligible because the energy gap to be crossed is very large compared to typical free electron energies (the only notable exceptions being collisional excitation from the \ion{H}{i} ground level in very hot objects \citep{1985ApJS...58..321D, 2003ApJ...592..846L} and of the \ion{He}{i} 2$^3$S metastable level in medium- to high-density objects); and the latter is negligible because of the weakness of the stellar ionizing spectra at the relevant energy values \citep{1999PASP..111.1524F, 2009ApJ...691.1712L}.

\subsection{Heavy elements: collisional lines}

In contrast to H and He, several astrophysically common metal ions have levels lying just a few eV above the ground level that be easily populated through collisions with free electrons. Such a mechanism is progressively less efficient for higher lying levels, so that only the lowest levels are appreciably populated through this mecshanism.  As a consequence, only the first few levels need to be considered, and the atom can be represented as a simple $n-$level system, with
the equilibrium level populations obeying the following set of equations:

\begin{eqnarray}
     \sum_{j\ne i} N_\mathrm{e} n_{j} q_{ji} & + & \sum_{j>i}  n_{j} A_{ji} = \nonumber \\
     \sum_{j\ne i} N_\mathrm{e} n_{i} q_{ij} & + & \sum_{j<i} n_{i} A_{ij}, \qquad (i = 1, ...n_{\mathrm max})  \label{eq:pop_balance}
\end{eqnarray}
and:
\begin{eqnarray}
    \sum_{i} n_{i} = n_\mathrm{tot},  \label{eq:pop_closure} 
\end{eqnarray}

\noindent where $N_\mathrm{e}$ is the electron density, 
$n_\mathrm{tot} $ the total density of the ion,
$n_{i}$ the density of ions with an electron on level $i$, 
$N_\mathrm{e} n_{j} q_{ji}$  is the rate of collisional excitation (if $j<i$) or de-excitation (if $j>i$), and $A_{ji}$ ($j>i$) is the transition probability from level $j$ to $i$.
Densities are measured in cm$^{-3}$ and transition probabilities in s$^{-1}$.
The number of levels considered $n$ is five or six for most ions, but the current database contains atoms with as few as two or as many as 34 levels. The rates of collisional de-excitation and excitation are derived from the effective collision strengths $\Upsilon$ through the following equations:

\begin{eqnarray*}
 q_{ji} &=& \frac{8.629 \cdot 10^{-6}}{g_{j}}\frac{\Upsilon(T_\mathrm{e})_{jk}}{T_\mathrm{e}^{1/2}},  \\ \label{eq:exc_rate}
\end{eqnarray*}
and
\begin{eqnarray*}
q_{ij} &= &\frac{g_{j}}{g_{i}}q_{ji}\; e^{-\Delta E_{ij}/k_bT_\mathrm{e}}, \\ \label{eq:de-exc_rate}
\end{eqnarray*}

\noindent where 
$j>i$, $g_{i}$, $g_{j}$, are the statistical weights of levels $i$ and $j$, respectively, and 
exp$({-\Delta E_{ij}/k_bT_\mathrm{e}})$ 
is the Boltzmann factor
\citep[see][and references therein]{2006agna.book.....O}.
The $\Upsilon$s are the result of atomic physics computations and are generally available in tabular form as a function of electron temperature.

\subsection{Plasma diagnostics}

Equations~\ref{eq:pop_balance} and  \ref{eq:pop_closure} represent a set of $n_\mathrm{tot}+1$ coupled equations in the unknowns $n_{j}$, which, for any given $N_\mathrm{e}, T_\mathrm{e}$ combination, can be solved for the relative level populations $n_{j}/n_\mathrm{tot}$.
Once the populations are known, the line emissivities are computed as

\begin{eqnarray}
 \epsilon_{ji} &=& n_{j} A_{ji} h\nu_{ji}.  
 \label{eq:emissivity}
\end{eqnarray}

A ratio of two lines emitted by the same ion depends on $N_\mathrm{e}$ and $T_\mathrm{e}$, but it does not depend on the total ion abundance $n_\mathrm{tot}$, which cancels out.
Since for cospatial ions the emissivity ratio is equal to the intensity ratio, an observed intensity ratio can be used to determine either of the two quantities given the other.

\subsection{Ionic and elemental abundances}\label{sec:abun}

Once the physical conditions are known, Eq.~\ref{eq:emissivity} can be manipulated to compute the ionic abundances of observed ions relative to H$^+$:

\begin{equation}
 \frac{n(\mathrm{X}^i)}{n(\mathrm{H}^+)} = \frac{I(\lambda)}{I(\rm{H}\beta)} \frac{\epsilon(\rm{H}\beta)}{\epsilon(\lambda)},  
\label{eq:ion_abun}
\end{equation}

\noindent where X$^i$ is an ion emitting a line of wavelength $\lambda$. 
If all the expected ions of a given element are observed, the elemental abundance is trivially given by the sum of the ionic abundances:

\begin{equation}
 \frac{n(\mathrm{X})}{n(\mathrm{H})} = \sum_{i}  \frac{n(\mathrm{X}^i)}{n(\mathrm{H}^+)}.
\label{eq:elem_abun}
\end{equation}

In most cases, however, not all relevant ions are observed and the incomplete summation must be corrected for unseen ions, using an ionization correction factor, or ICF: 

\begin{equation}
\frac{n(\mathrm{X})}{n(\mathrm{H})} = \sum_{\bar{\imath}} \frac{n(\mathrm{X}^{\bar{\imath}})}{n(\mathrm{H}^+)} \times \mathrm{ICF},
\label{eq:icf}
\end{equation}

\noindent where the index ${\bar{\imath}}$ now extends only over a subset of ions. 

The ICFs incorporate some form of knowledge or expectation on the ionization structure of the nebula, based either on the results of photoionization models or on comparisons between the ionization potentials of different ions. In the literature one can find many such expressions, most of which have been devised for a specific kind of object (e.g., PNe, \ion{H}{ii} regions, etc.) and should therefore not be applied to objects of a different kind.

\section{Overview of the code}\label{sec:overview}

Fundamentally, PyNeb is a toolbox meant to enable nebular analysis using direct methods and to visualize the results. Its key functionality is to solve for line emissivities, determine electron temperature and density given observed diagnostic  line ratios, and compute (within some simplifying assumptions) ionic and total abundances for a large number of elements that are typically observed in gaseous nebulae. PyNeb offers a variety of methods, visualization tools, and programming abstractions to make this easy for a user. It is also designed to be easy for end-users with modest knowledge of scientific programming and the python language to update, extend, and customize PyNeb for particular science objectives by, for example, building scripts to determine abundances from observed emission lines, using the core PyNeb classes to build new software, or developing a new user interface. Users can also add new or alternate atomic data. The list of currently supported ions is given in Table~\ref{tab:ion_support}, as well as online. 

The key concepts when using PyNeb are the internal model of the ion, which is embodied in the \textbf{Atom} and the \textbf{RecAtom} classes; the representation of a set of observed UV/Optical/IR emission lines, which is embodied in the \textbf{Observation} class; the representation of the plasma diagnostic ratios, embodied in the \textbf{Diagnostics} class; and the representation of ionization correction factors, embodied in the \textbf{ICF} class. 
The available methods for these classes, plus some ancillary classes, are the means by which the data can be visualized and the analysis carried out. The methods are intended to support a variety of workflows, from simple computations of diagnostics to visualizing  the supporting atomic data, to complete abundance analyses. 
Beyond that there is a \textbf{Stellar} class for the special case of computing the effective temperature of a PN central star (via the Zanstra technique), as well as its luminosity. 

\begin{table}
\centering
      \caption[]{Supported ions in PyNeb\label{tab:ion_support}}
\begin{tabular}{cccccc}
\hline \hline
Atom		& Ion		& Ground state & Atom		& Ion		& Ground state\\
\hline
Al 	& II	& $s^2$ &   N   	& I	    & $p^3$	\\	
Ar 	& II	& $p^5$ &   		& II	& $p^2$	\\
	& III	& $p^4$ &   		& III	& $p^1$	\\
	& IV	& $p^3$ &   		& IV	& $s^2$	\\
	& V	    & $p^2$ &   	Na 	& IV	& $p^4$	\\
Ba 	& II	& $s^1$ &   	    & VI	& $p^2$	\\
 	& IV	& $p^5$ &   	Ne	& II	& $p^5$	\\
Br 	& III	& $p^3$ &   	    & III	& $p^4$	\\
 	& IV	& $p^2$ &   		& IV	& $p^3$	\\
C 	& I  	& $p^2$ &   		& V	    & $p^2$	\\
 	& II	& $p^1$ &   		& VI	& $p^1$	\\
 	& III	& $s^2$ &   	O 	& I	    & $p^4$	\\
 	& IV	& $s^1$ &   		& II	& $p^3$	\\
Ca 	& V	    & $p^4$	&		   	& III	& $p^2$	\\
Cl 	& II	& $p^4$ &   		& IV	& $p^1$	\\
	& III	& $p^3$ &   	    & V	    & $s^2$	\\
	& IV	& $p^2$ &   	Rb 	& IV	& $p^4$	\\
Fe	& III	& $s^2$ &   		& V  	& $p^3$	\\
K 	& IV	& $p^4$ &   	S	& II	& $p^3$	\\
	& V	    & $p^3$ &   		& III	& $p^2$	\\
Kr 	& III	& $p^4$ &   		& IV	& $p^1$	\\
	& IV	& $p^3$ &   	Se	& III	& $p^2$	\\
	& V	    & $p^2$ &   		& IV	& $p^1$	\\
Mg 	& V	    & $p^4$ &   	Si	& II	& $p^1$	\\
	& VII	& $p^2$ &   		& III	& $s^2$	\\
Xe 	& III	& $p^4$	&&& \\
	& IV	& $p^3$	&&& \\

\hline
\end{tabular}
\end{table}

New users may find studying the example scripts (available from the web page or on request through the forum; see Sect.~\ref{sec:documentation}) very helpful for worked examples of PyNeb capabilities. Each script may be executed from the host command line or imported into an active python session. The following sections provide a brief tour and highlight important scientific functionality, as well as details of key methods and their output.

\subsection{PyNeb core functionality}\label{sec:core_functionality}

\begin{figure*}
   \centering
  \includegraphics[width=18cm]{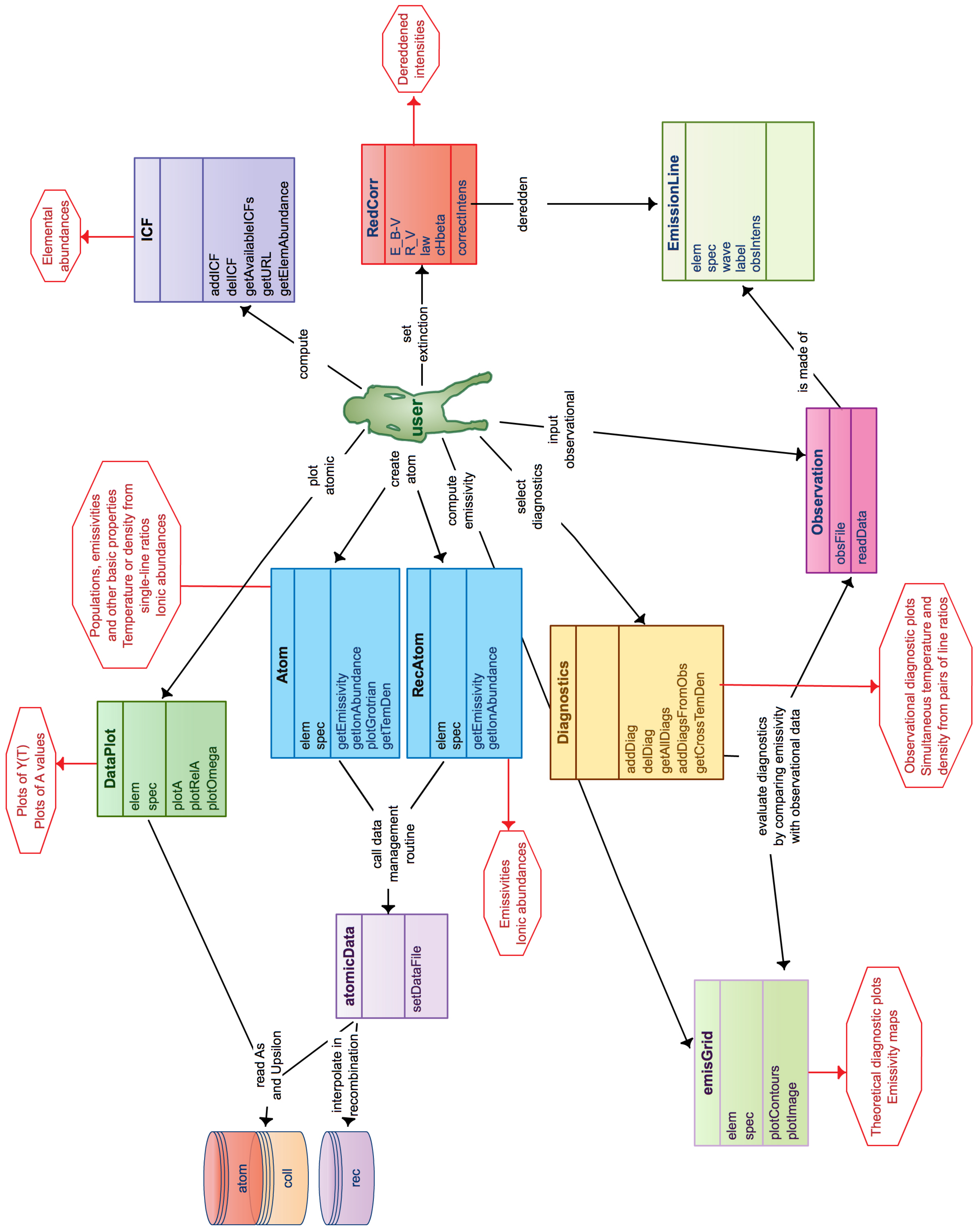}
   \caption{Schematic view of PyNeb's structure in terms of its classes (rectangular boxes) and output quantities (red octagons). The first slot of each box contains the class name, the second the class arguments, and the third the class methods. The description is abridged and only meant for illustrative purposes.} 
              \label{fig:sketch}%
    \end{figure*}

The main PyNeb's classes and methods are represented in Fig.~\ref{fig:sketch}, which also schematically describes how they interact with each other and the user. 
In the following sections, we describe the scope of the main classes. The syntax will not be described in detail, since it is fully treated in the code documentation (see Sect.~\ref{sec:documentation}).

PyNeb distinguishes between atoms emitting collisionally-excited lines (called here "collisional atoms") and atoms emitting recombination lines ("recombination atoms").
Both are represented as {\sl n}-level systems and share the same basic syntax, but they are encoded in different classes and their internal representation is different in either case, as detailed in the following sections.

\subsection{The model collisional atom: The Atom class}\label{sec:Atom}

The critical features of the physical atom relevant for our scope are encapsulated, for the collisional case, in the \texttt{Atom} data structure. For example, the following command, defining an [O{\sc~iii}] atom:

\begin{verbatim}
o3 = pn.Atom('O', 3)
\end{verbatim}

\noindent creates a data structure capable of holding the $A$ and $\Upsilon$ arrays and fills it with the actual [O{\sc~iii}] data read from the database. The resulting object is also equipped with the calculus machinery necessary to: solve the system of Eqs.\ref{eq:pop_balance} and \ref{eq:pop_closure} to compute the level populations; apply Eq.~\ref{eq:emissivity} to calculate the emissivities; derive the ion abundance from Eq.~\ref{eq:ion_abun}; and compute or output several other standard quantities.  

From the above, it follows that the number of levels of a particular ion is not intrinsic to the ion but instead depends on the data used. For most ions, different data sets are available, corresponding to different atomic data calculations, and changing the data set may possibly imply a change in the number of levels and, almost inevitably, in the result of calculations too. 

One asset of PyNeb over similar programs is that all the atomic properties computed by 
PyNeb can be retrieved through manipulations of this Atom object; for example:

\begin{verbatim}
o3.getPopulations(tem, den)
o3.getCritDensity(tem, den)
\end{verbatim}

\noindent return the populations and critical densities, respectively, at the specified physical conditions $T_\mathrm{e}=tem$ and $N_\mathrm{e}=den$.
Properties made accessible by PyNeb include the name, the spectrum, the level energies,  the transition probabilities, the set of $\Upsilon$s at a specified temperature, the Grotrian diagram, and many others. The complete inventory can be displayed and explored using standard python commands.

Atom objects are also equipped with basic functionality for determining physical conditions, i.e. the \texttt{Atom.getTemDen} method:

\begin{verbatim}
o3.getTemDen(int_ratio=0.01, den=1000,  
             wave1=4363, wave2=5007),
\end{verbatim} 

\noindent which returns the electronic temperature such that [O{\sc~iii}] $\lambda$4363/$\lambda$5007=0.01 at $N_\mathrm{e}$=1000 cm$^{-3}$.

\subsection{Atoms emitting recombination lines: the RecAtom class}\label{sec:RecAtom}

In the case of recombination lines, PyNeb computes the emissivity of a given line by either interpolating in tables or using a fitting function. This is accomplished by the \texttt{RecAtom class}: for example, the H{\sc~i} atom (or, to be more precise, the recombination spectrum of H{\sc~i}) is instantiated through the command

\begin{verbatim} 
h1=pn.RecAtom(elem='H', spec=1). 
\end{verbatim} 

Recombination atoms share many of the features of collisional atoms and follow the same syntax: for example, the \texttt{h1} object defined above can be explored with commands such as

\begin{verbatim}
h1.name
h1.getEmissivity(tem=10000, den=1000, label='4_2'),
\end{verbatim}

\noindent which are the recombination counterparts of the analogous methods of the Atom class.

If both the recombination and the collisional spectrum of the same atom are needed, two different objects must be built; thus, \texttt{pn.Atom('O', 2)} would describe the behavior of collisionally excited lines of O$^{+}$ –the [O{\sc~ii}] spectrum–, while \texttt{pn.RecAtom('O', 2)} would describe the behavior of recombination lines of the same ion –the O{\sc~ii} spectrum.

\subsection{Determining ionic and total abundances with PyNeb} 
 
If the physical conditions are known, it is straightforward to use the observed line intensities to determine the ionic abundance of any given $X^i$ ion relative to H$^+$ by using Eq.~\ref{eq:ion_abun}, which, in PyNeb, is implemented as the \texttt{Atom} method \texttt{getIonAbundance}:

\begin{verbatim}
o3.getIonAbundance(int_ratio=100, tem=1.5e4, \
                   den=100, wave=5007).
\end{verbatim}

Total abundances are computed by PyNeb following the ICF formalism described in Sect.~\ref{sec:abun}.
The code includes a large inventory of ICFs and a series of methods to explore them and their source papers.
Table~\ref{tab:icf} lists the published ICFs included in version 1.0.1 of PyNeb. 
More expressions are being added constantly. PyNeb also features a method (\texttt{addICF}) to add customized expressions to the collection.   

   \begin{table}
\centering
      \caption[]{ICFs included in version 1.0.1. of PyNeb}
         \label{tab:icf}
\begin{tabular}{llc}
\hline \hline
Source      &  Elements  & Type \\
\hline
  {\protect\citet{2014MNRAS.440..536D}}	& He, O, N, Ne &  \\             
   & S, Cl, Ar, C & PNe \\             
  {\protect\citet{2007A&A...463..265G}}	& Cl & PNe \\             
  {\protect\citet{2006A&A...448..955I}}	& N, O, Ne, S, & \\
  & Cl, Ar, Fe& H II\\
  {\protect\citet{1994MNRAS.271..257K}}	& C, N, O, Ne, &  \\
  & S, Ar & PNe\\
  {\protect\citet{2001ApJ...562..804K}}	& He, N, O, &  \\
  & Ne, Cl, Ar& PNe\\
  {\protect\citet{1969BOTT....5....3P}}		& Ne & H II \\
  {\protect\citet{1992RMxAA..24..155P}}	& He & H II \\
  {\protect\citet{2007MNRAS.381..125P}} & Ne, Ar & H II \\
  {\protect\citet{2005ApJ...626..900R}}	& Fe & H II \\
  {\protect\citet{1978A&A....66..257S}} 	& Ne & H II \\
  {\protect\citet{1977RMxAA...2..181T}}	& N, O, Ne & PNe \\
\hline
\end{tabular}
\end{table}

\subsection{Physical condition diagnostics}\label{sec:diagnostics}

The simple diagnostic calculation described in Sect.~\ref{sec:Atom}, which is a method of the \texttt{Atom} class, only involves a line ratio. The analysis of real objects often requires several diagnostic ratios to be considered simultaneously; doing this requires the use of a dedicated class, \texttt{Diagnostics}. A \texttt{Diagnostics} object, which is created by the command

\begin{verbatim}
diags=pn.Diagnostics(),
\end{verbatim}

\noindent is a container for all the diagnostic ratios used during a calculation. Each diagnostic is associated to a label, the symbol and spectrum of the emitting ion, the defining expression in term of individual lines, and the rms error as a function of the rms error of the included lines. 

   \begin{table*}
\centering 
      \caption[]{Default line ratios included in version 1.0.1 of PyNeb. The suffix '+' indicates that the line is a blend.} 
      \label{tab:diags} 
\begin{tabular}{lllll}
\hline 
\hline 
{[}Ar{\sc~iii}] $\lambda$5192/$\lambda$7136   & Fe{\sc~iii}] $\lambda$4735/$\lambda$4009               & [Fe{\sc~iii}] $\lambda$4987/$\lambda$4703 & [N{\sc~i}] $\lambda$5198/$\lambda$5200               & [O{\sc~iii}] $\lambda$1666/$\lambda$4363     \\
{[}Ar{\sc~iii}] $\lambda$5192/$\lambda$7300+  & Fe{\sc~iii}] $\lambda$4882/$\lambda$4009               & [Fe{\sc~iii}] $\lambda$4987/$\lambda$4735 & [N{\sc~ii}] $\lambda$121$\mu$/$\lambda$20.5$\mu$     & [O{\sc~iii}] $\lambda$1666/$\lambda$5007     \\
{[}Ar{\sc~iii}] $\lambda$7136/$\lambda$9$\mu$ & Fe{\sc~iii}] $\lambda$4882/$\lambda$4659               & [Fe{\sc~iii}] $\lambda$4987/$\lambda$4882 & [N{\sc~ii}] $\lambda$5755/$\lambda$6548              & [O{\sc~iii}] $\lambda$1666/$\lambda$5007+    \\
{[}Ar{\sc~iii}] $\lambda\lambda$7751, 7136/$\lambda$9$\mu$ & Fe{\sc~iii}] $\lambda$4882/$\lambda$4703  & [Fe{\sc~iii}] $\lambda$4987/$\lambda$4926 & [N{\sc~ii}] $\lambda$5755/$\lambda$6584              & [O{\sc~iii}] $\lambda$4363/$\lambda$5007     \\
{[}Ar{\sc~iii}] $\lambda$9.0$\mu$/$\lambda$21.8$\mu$ & [Fe{\sc~iii}] $\lambda$4882/$\lambda$4932       & [Fe{\sc~iii}] $\lambda$4987/$\lambda$4932 & [N{\sc~ii}] $\lambda$5755/$\lambda$6584+             & [O{\sc~iii}] $\lambda$4363/$\lambda$5007+    \\
{[}Ar{\sc~iv}] $\lambda$2860+/$\lambda$4720+  & [Fe{\sc~iii}] $\lambda$4882/$\lambda$5013              & [Fe{\sc~iii}] $\lambda$4987/$\lambda$5013 & [Ne{\sc~iii}] $\lambda$15.6$\mu$/$\lambda$36.0$\mu$  & [O{\sc~iii}] $\lambda$5007/$\lambda$88$\mu$  \\
{[}Ar{\sc~iv}] $\lambda$4740/$\lambda$4711    & [Fe{\sc~iii}] $\lambda$4926/$\lambda$4009              & [Fe{\sc~iii}] $\lambda$5013/$\lambda$4009 & [N{\sc~iii}] $\lambda$1750/$\lambda$57.4$\mu$        & [O{\sc~iii}] $\lambda$51$\mu$/$\lambda$88$\mu$     \\
{[}Ar{\sc~iv}] $\lambda$7230+/$\lambda$4720+  & [Fe{\sc~iii}] $\lambda$4926/$\lambda$4659              & [Fe{\sc~iii}] $\lambda$5013/$\lambda$4659 & [Ne{\sc~iii}] $\lambda$3343/$\lambda$3930+           & [O{\sc~iv}] $\lambda$1401/$\lambda$1405            \\
{[}Ar{\sc~v}] $\lambda$4626/$\lambda$6600+    & [Fe{\sc~iii}] $\lambda$4926/$\lambda$4703              & [Fe{\sc~iii}] $\lambda$5013/$\lambda$4735 & [Ne{\sc~iii}] $\lambda$3344/$\lambda$3930+           & [S{\sc~iii}] $\lambda$18.7$\mu$/$\lambda$33.6$\mu$ \\
{[}C{\sc~iii}] $\lambda$1909/$\lambda$1907    & [Fe{\sc~iii}] $\lambda$4926/$\lambda$4735              & [Fe{\sc~iii}] $\lambda$5013/$\lambda$4932 & [Ne{\sc~iii}] $\lambda$3869/$\lambda$15.6$\mu$       & [S{\sc~iii}] $\lambda$6312$\lambda$/18.7$\mu$      \\
{[}Cl{\sc~iii}] $\lambda$5538/$\lambda$5518   & [Fe{\sc~iii}] $\lambda$4926/$\lambda$4882              & [Fe{\sc~iii}] $\lambda$5272/$\lambda$4009 & [Ne{\sc~iii}] $\lambda$3930+/$\lambda$15.6$\mu$      & [S{\sc~iii}] $\lambda$6312/$\lambda$9069           \\ 
{[}Cl{\sc~iv}] $\lambda$5323/$\lambda$7531    & [Fe{\sc~iii}] $\lambda$4926/$\lambda$4932              & [Fe{\sc~iii}] $\lambda$5272/$\lambda$4659 & [Ne{\sc~v}] $\lambda$14.3$\mu$/$\lambda$24.2$\mu$    & [S{\sc~iii}] $\lambda$6312/$\lambda$9200+          \\ 
{[}Cl{\sc~iv}] $\lambda$5323/$\lambda$7700+   & [Fe{\sc~iii}] $\lambda$4926/$\lambda$5013              & [Fe{\sc~iii}] $\lambda$5272/$\lambda$4703 & [Ne{\sc~v}] $\lambda$2975/$\lambda$3370+             & [S{\sc~iii}] $\lambda$9069/$\lambda$18.7$\mu$      \\
{[}Fe{\sc~iii}] $\lambda$4659/$\lambda$4009   & [Fe{\sc~iii}] $\lambda$4932/$\lambda$4009              & [Fe{\sc~iii}] $\lambda$5272/$\lambda$4735 & [O{\sc~i}] $\lambda$5579/$\lambda$6300               & [S{\sc~ii}] $\lambda$4069/$\lambda$4076            \\
{[}Fe{\sc~iii}] $\lambda$4659/$\lambda$4703   & [Fe{\sc~iii}] $\lambda$4932/$\lambda$4659              & [Fe{\sc~iii}] $\lambda$5272/$\lambda$4882 & [O{\sc~i}] $\lambda$5579/$\lambda$6300+              & [S{\sc~ii}] $\lambda$4072+/$\lambda$6720+          \\
{[}Fe{\sc~iii}] $\lambda$4659/$\lambda$4735   & [Fe{\sc~iii}] $\lambda$4932/$\lambda$4703              & [Fe{\sc~iii}] $\lambda$5272/$\lambda$4926 & [O{\sc~i}] $\lambda$5579/$\lambda$6302               & [S{\sc~ii}] $\lambda$6731/$\lambda$6716            \\
{[}Fe{\sc~iii}] $\lambda$4659+/$\lambda$4987+ & [Fe{\sc~iii}] $\lambda$4932/$\lambda$4735              & [Fe{\sc~iii}] $\lambda$5272/$\lambda$4932 & [O{\sc~i}] $\lambda$63$\mu$/$\lambda$147$\mu$   & \\
{[}Fe{\sc~iii}] $\lambda$4703/$\lambda$4009   & [Fe{\sc~iii}] $\lambda$4987/$\lambda$4009              & [Fe{\sc~iii}] $\lambda$5272/$\lambda$4987 & [O{\sc~ii}] $\lambda$3726/$\lambda$3729         & \\
{[}Fe{\sc~iii}] $\lambda$4703/$\lambda$4735   & [Fe{\sc~iii}] $\lambda$4987/$\lambda$4659              & [Fe{\sc~iii}] $\lambda$5272/$\lambda$5013 & [O{\sc~ii}] $\lambda$3727+/$\lambda$7325+       & \\
\hline 
   \end{tabular}
      \end{table*}

A large number of published diagnostic ratios are stored in PyNeb by default and other ones are continuously added. Those included in the code at the time of writing this paper are shown in Table~\ref{tab:diags}. 
User-defined diagnostics can also be added to PyNeb's built-in repertoire by adding any combination of lines of the same ion, but it is the user's responsibility to ensure that the diagnostics added are meaningful and that the correct expression for the rms is given.

The most important feature of the \texttt{Diagnostics} class is the \texttt{getCrossTemDen method}, which allows simultaneously determining the temperature and density by fitting two line ratios:

\small
\begin{verbatim}
tem, den = diags.getCrossTemDen(
                       diag_tem='[NII] 5755/6548',
                       diag_den='[SII] 6731/6716',
                       value_tem=50,value_den=1.0).
\end{verbatim}
\normalsize

\noindent In this command, the first two parameters are the labels of two line ratios and the following two are the corresponding values. The command stores the resulting $T_\mathrm{e}$ and $N_\mathrm{e}$ values in the variables \texttt{tem} and \texttt{den}, which can be used in subsequent calculations.
This method works by making an initial guess of the temperature, after which it applies a simple multisection procedure that iteratively calls getTemDen, until a solution is found within a given precision.

\subsection{Emissivities as a function of temperature and density: EmisGrid}

\begin{figure*}
   \centering
      \includegraphics[width=15cm,clip]{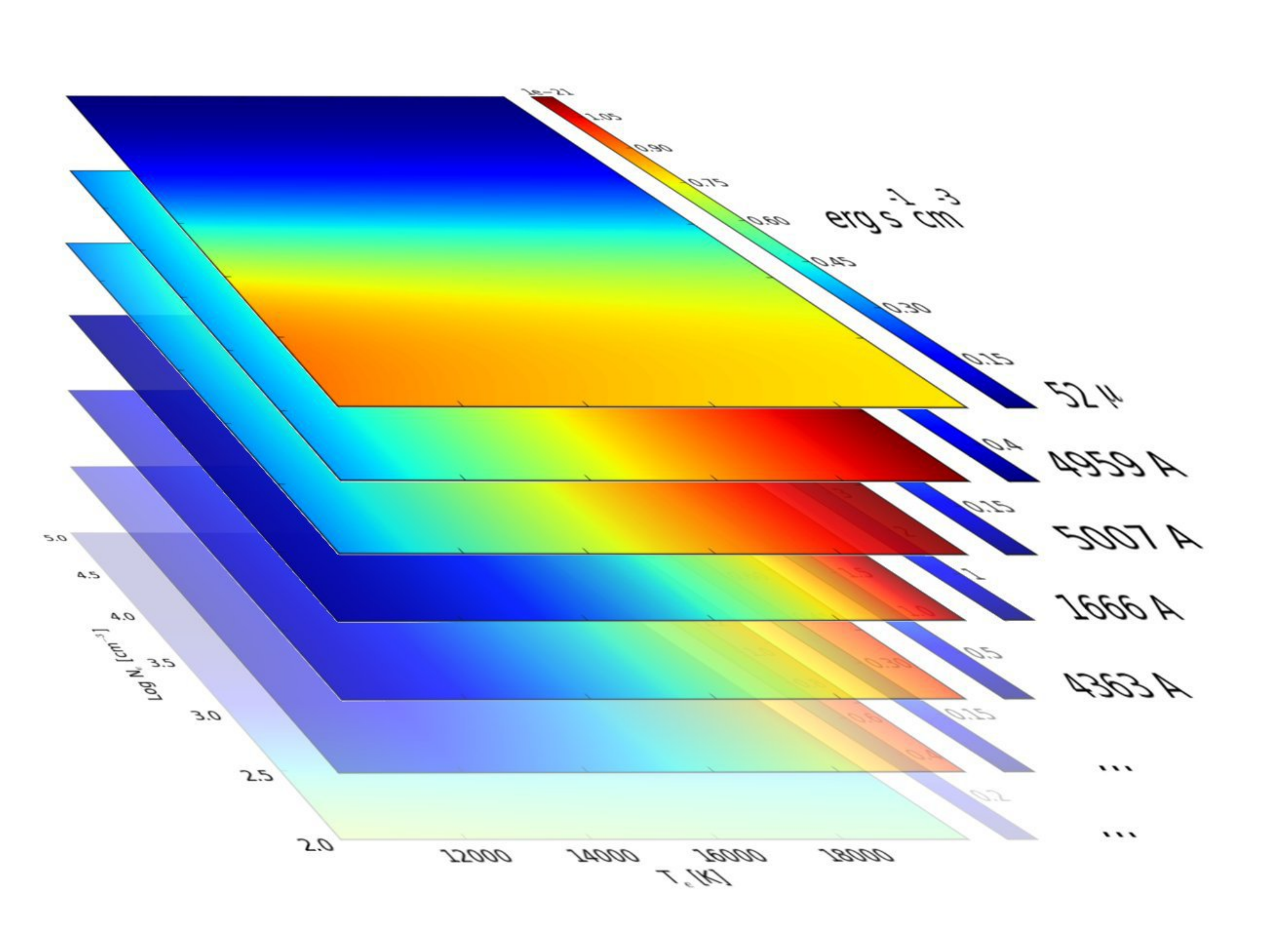}
      \caption{EmisGrid object for O{\sc~iii}. Each layer in the stack represents the emissivity of an O{\sc~iii} line in the chosen $T_\mathrm{e}$, $N_\mathrm{e}$ range.}
         \label{fig:emisgrid}
   \end{figure*}

\begin{figure*}
   \centering
      \subfigure[]{%
            \label{fig:first}
           \includegraphics[width=0.38\textwidth]{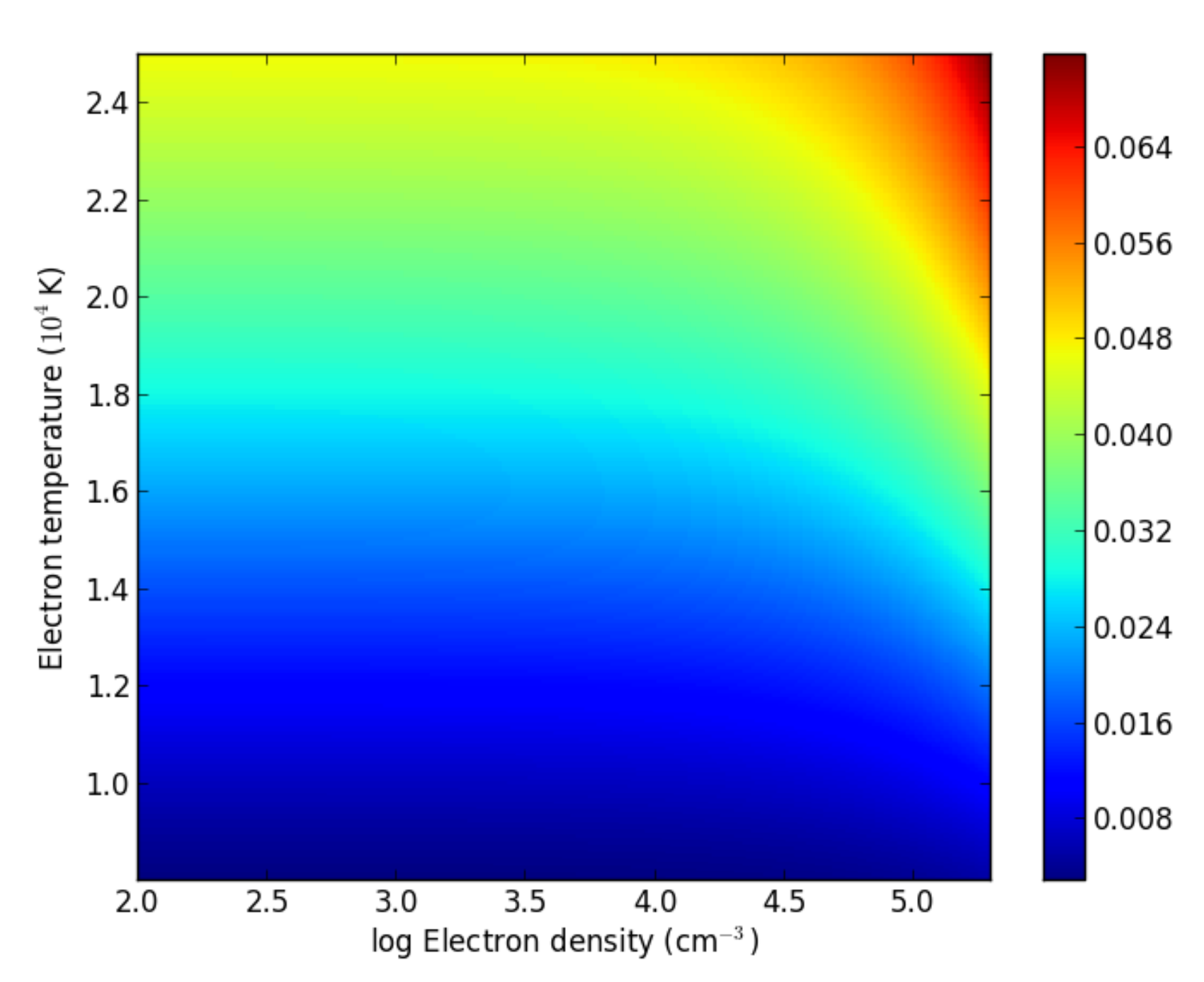}
        }%
      \subfigure[]{%
            \label{fig:second}
              \includegraphics[width=0.43\textwidth]{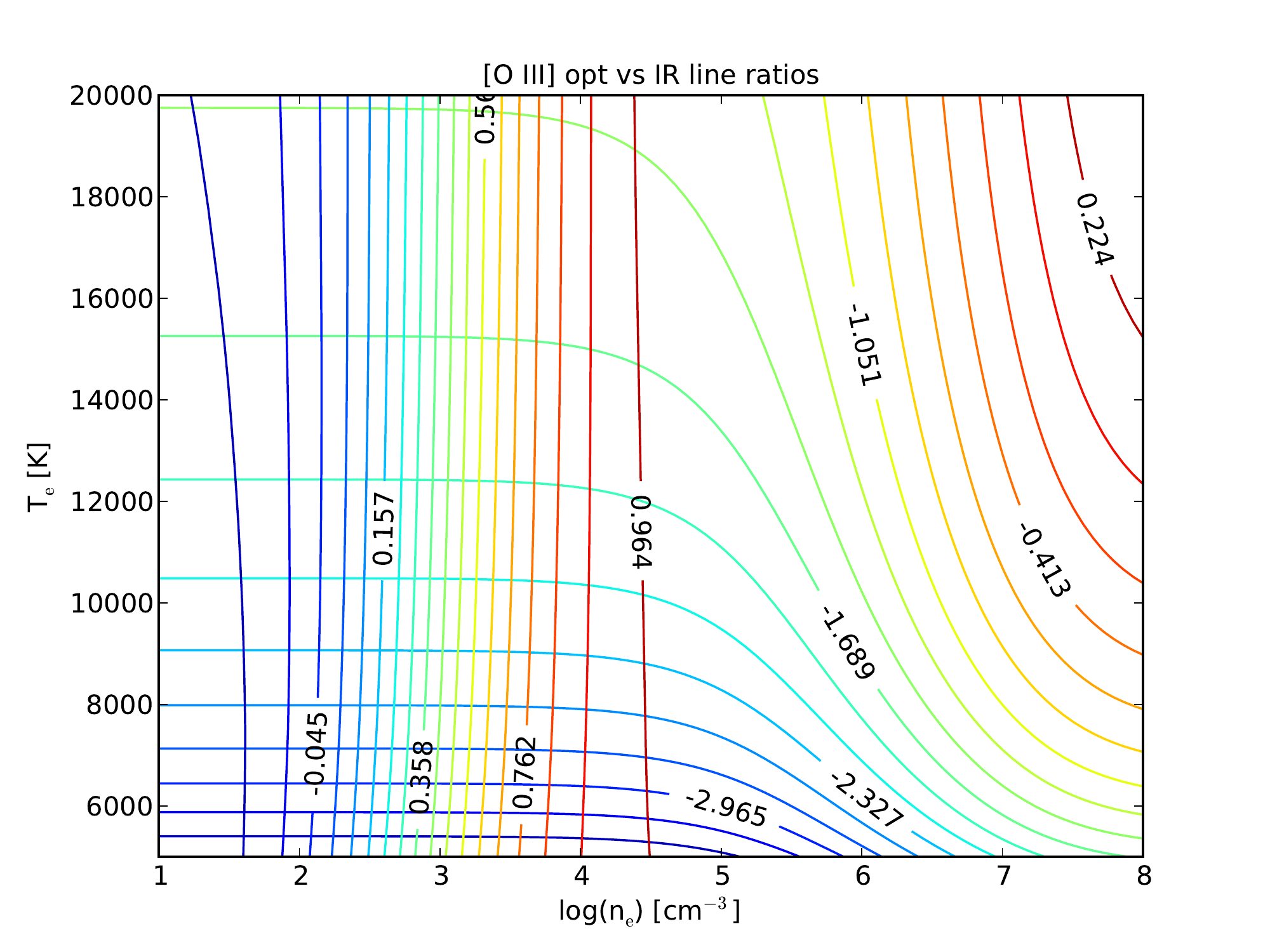}
              }
   \caption{Left: map of the [O{\sc~iii}] $\lambda$4363/$\lambda$5007 intensity ratio as a function of electron density and temperature, obtained with the plotImage() command. Right: theoretical contour plot for the simultaneous determination of $T_\mathrm{e}$ and $N_\mathrm{e}$ with the [O{\sc~iii}] $\lambda\lambda 4959, 5007$/$\lambda 4363$ (mainly horizontal curves) and [O{\sc~iii}] $\lambda 52\mu$/$\lambda 88\mu$ (vertical curves) intensity ratios as a function of electronic density and temperature, obtained with the plotContours() command. The levels are numbered with the log of the ratio.}
              \label{fig:emisgrid-plots}%
   \end{figure*}

Many of PyNeb's tasks are executed by manipulating precalculated grids of line emissivities as a function of temperature and density, obtained by instantiating the class \texttt{EmisGrid}, which produces a stack of emission grids for the selected atom. A stack has as many layers as lines in the atom (see the example of the [O{\sc~iii}] \texttt{EmisGrid} object in Fig.~\ref{fig:emisgrid}). The class is equipped with methods to display the information in various ways, such as color-coded emissivity maps and contour plots (Fig.~\ref{fig:emisgrid-plots}). 

%

\subsection{Managing observational data: the Observation class}\label{sec:observations} 
  
\begin{figure*}
   \centering
   \includegraphics[width=12cm]{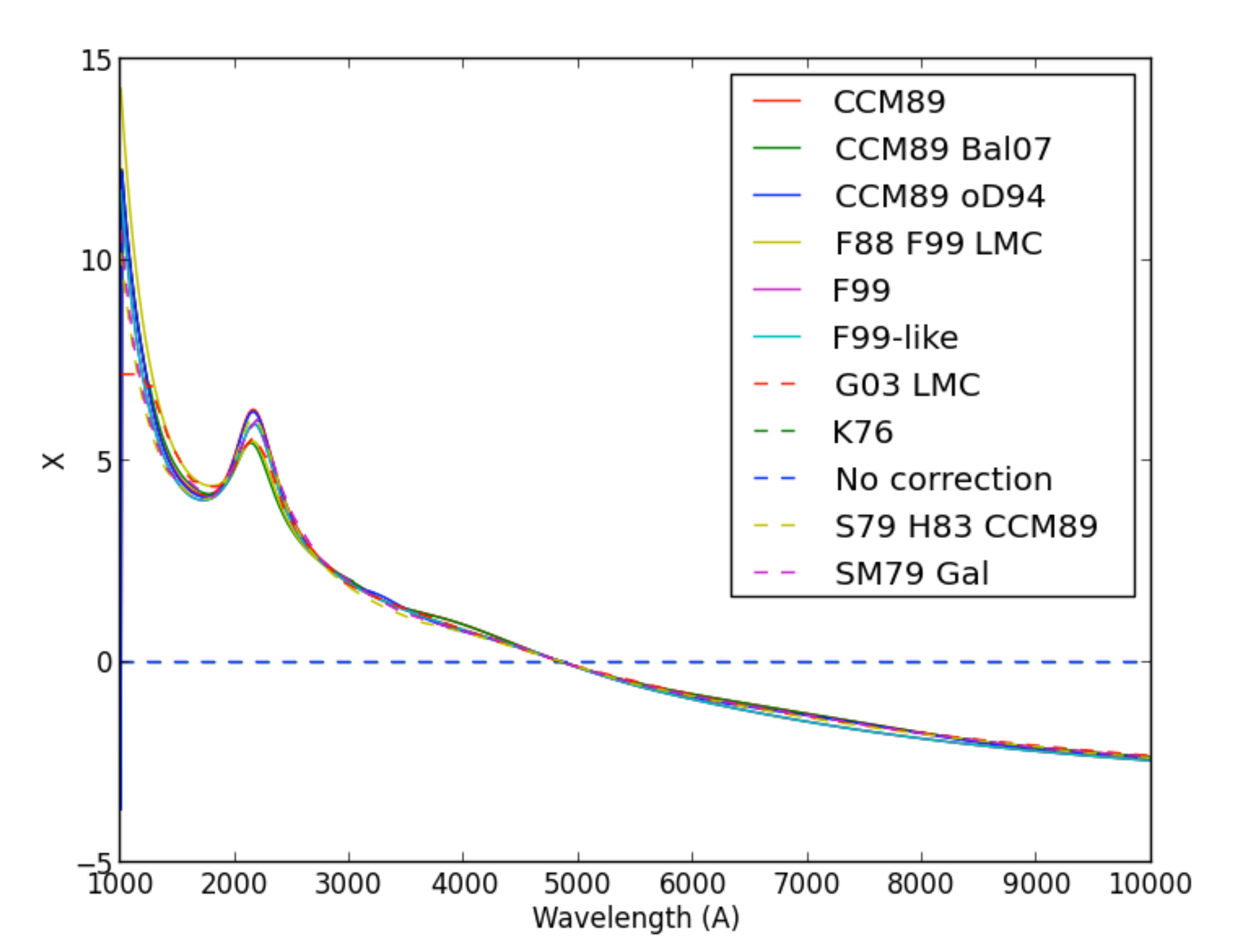}
   \caption{Plot of the built-in extinction laws included in PyNeb. Details on the laws can be displayed by typing \texttt{pn.RedCorr().printLaws()}. Additional laws can be provided by the user, either in tabular form or as an analytical formula. The plot has been obtained with the \texttt{RC.plot(laws='all')} command, where \texttt{RC} is an instance of \texttt{pn.RedCorr()}.}
              \label{fig:ext_laws}%
\end{figure*}

Pyneb's \texttt{Observation} class creates the data structure in charge of holding observational data. An \texttt{Observation} object includes at least a set of line intensities for one or more objects and may include other related information, such as observational errors, the $c($H$\beta)$ coefficient, the dereddened line intensities, and an indication of the dereddening law used (which may be either built-in or user-defined; see Table~\ref{tab:ext_laws} and Fig.~\ref{fig:ext_laws}). The data are generally read from a file: 
 
\begin{verbatim}
obs = pn.Observation()
obs.readData(obsFile='ic418.dat').
\end{verbatim}

\begin{table*}
\centering
      \caption[]{Extinction laws included in version 1.0.1 of PyNeb}
         \label{tab:ext_laws}
\begin{tabular}{llll}
\hline \hline
            Source      &  Label &  Type \\
            \hline
             {\protect\citet{1989ApJ...345..245C}} & 'CCM 89' & Galactic \\
             {\protect\citet{2007ApJ...655..299B}}, {\protect\citet{1989ApJ...345..245C}}  & 'CCM89 Bal07' & Galactic  \\
             {\protect\citet{1994ApJ...422..158O}}, {\protect\citet{1989ApJ...345..245C}} & 'CCM89 oD94' &  Galactic  \\
             {\protect\citet{1988ApJ...328..734F}}, {\protect\citet{1999PASP..111...63F}} & 'F88 F99 LMC' & LMC \\
             {\protect\citet{1999PASP..111...63F}} & 'F99' & Galactic \\
             {\protect\citet{1999PASP..111...63F}} & 'F99-like' &  \\
             {\protect\citet{2003ApJ...594..279G}} & 'G03 LMC' & LMC \\
             {\protect\citet{1976ApJS...31..517K}} & 'K76' & Galactic \\             
             {\protect\citet{1979MNRAS.187P..73S}}, {\protect\citet{1983MNRAS.203..301H}}, {\protect\citet{1989ApJ...345..245C}} & 'S79 H83 CCM89' & Galactic \\
             {\protect\citet{1979ARA&A..17...73S}} & 'SM79 Gal' &  Galactic \\             
            \hline
         \end{tabular}
   \end{table*}

\subsubsection{Observational diagnostics plots}\label{sec:obs_diags} 
  
\begin{figure*}
   \centering
  \includegraphics[width=12cm]{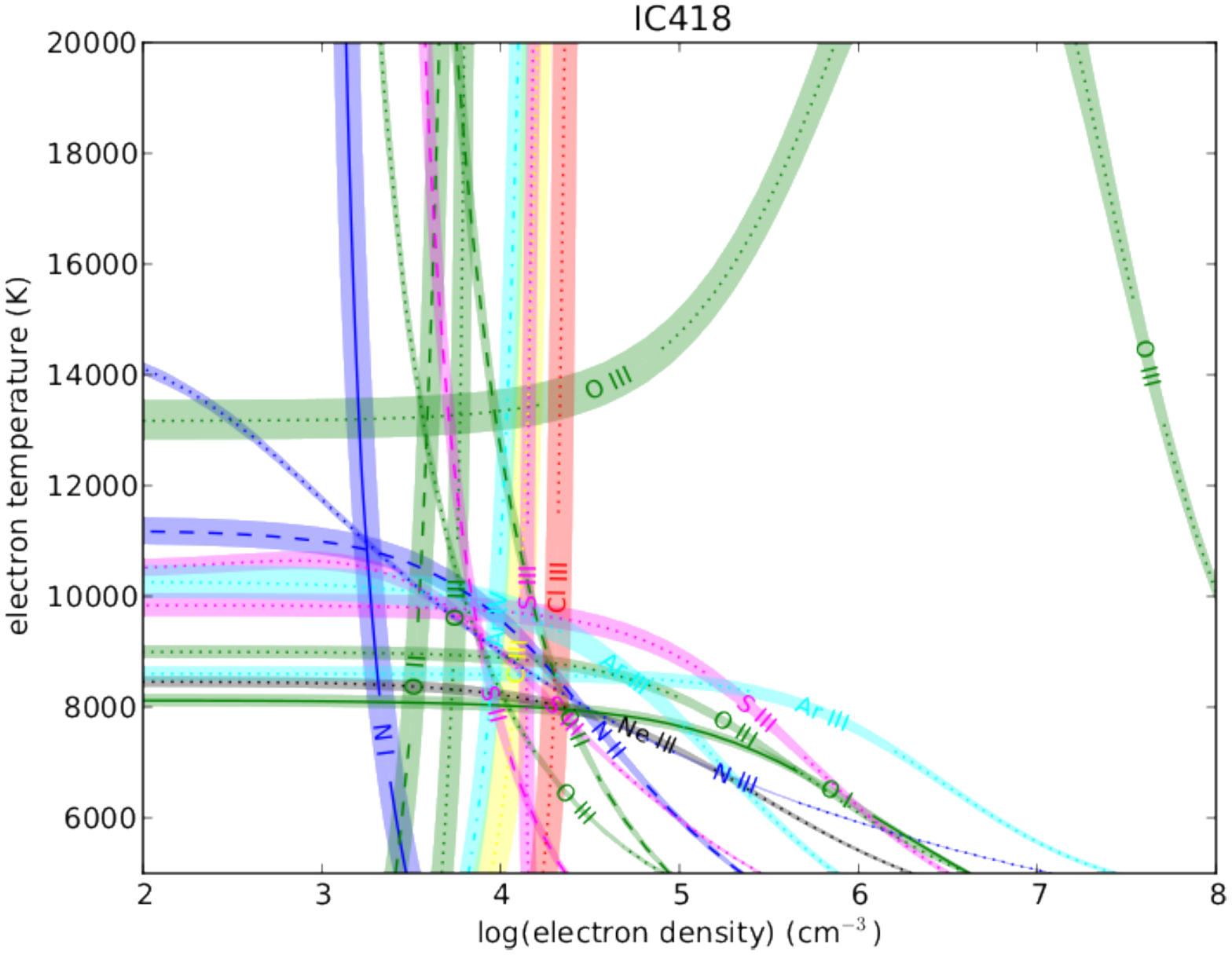}
   \caption{Emission-line diagnostic plot of the planetary nebula \object{IC 418}. Each band represents the $N_\mathrm{e}, T_\mathrm{e}$ locus consistent with the observed ratio and its uncertainty.  The observational data have been taken from the compilation in \citet{2009A&A...507.1517M} and an analysis of the object is given in \citet{2012IAUS..283..350E}. A sample script that produces this kind of diagnostic plots is given in the Appendix.}
              \label{fig:ic418}%
\end{figure*}

The joint use of Observation, Diagnostic, and EmisGrid objects makes it possible
to build observational diagnostic plots (Fig.~\ref{fig:ic418}).
We note that, when more than two lines ratios are available for a given object, a unique solution generally does not exist. The analysis of these plots requires taking observational errors into account and full consideration of all possible causes  
for the occurrence of "discordant" diagnostics in such plot, such as observational ones (e.g., biased or otherwise low-quality observational data), theoretical ones (e.g., emission grids computed with biased atomic data, unaccounted-for processes such as fluorescence, etc.) or physical ones (e.g., a complex ionization structure, which hinders direct comparison of line ratios of different ions). A full discussion of the specific case of fluorescent emission in \object{IC~418} is given in \citet{2012IAUS..283..350E} (see also Sect.~\ref{sec:limitations}).

\subsection{Atomic data} \label{sec:atomic_data}

As already stated in Section~\ref{sec:Atom}, the results of the calculations carried out by PyNeb depend on the assumed transition probabilities and collision strengths. These quantities are the result of 
complex atomic physics calculations and are subject to uncertainties. While the uncertainty is relatively low for some atoms, other atoms are more complex to model and the resulting uncertainty is correspondingly larger; in such cases, the result of nebular analysis may depend strongly on the assumed atomic data set. To address this issue, PyNeb provides a large data base with as many determinations as possible for each ion and a number of tools to assess the differences between them. The atomic data base and the related tools are an integral part of PyNeb's philosophy and will be described in Paper II of this series.

\section{Code validation}\label{sec:code_validation}

As mentioned above, PyNeb shares much of NEBULAR functionality; in particular, both are able to compute critical densities, level populations, line emissivities, electron temperatures and densities (given the other quantity and a line ratio),
and ionic abundances. Table~\ref{tab:tasks} lists the NEBULAR tasks involved and the corresponding PyNeb methods. The tests conducted show that PyNeb predictions coincide with those by NEBULAR (if, of course, the same atomic data are adopted). 
For other quantities computed by PyNeb, a direct comparison is not possible, 
either because NEBULAR does not compute them (as for elemental abundances) or because it does not output them (as for collisional rates);
but, even in the latter case, the consistency of the results clearly indicates that PyNeb is working as expected.

   \begin{table*}
\centering 
      \caption[]{Correspondence between NEBULAR's tasks and PyNeb's methods.} \label{tab:tasks} 
\begin{tabular}{lll}
\hline \hline 
            {NEBULAR}  &  PyNeb& Quantity computed \\
\hline 
ionic & getIonAbundance & Ionic abundances\\
ionic & printIonic & Critical densities\\
ionic & printIonic & Emissivities \\
ionic & Atom.printIonic, & H$\beta$ emissivity\\
      & RecAtom.getEmissivity &  \\
temden & getTemDen & $T_\mathrm{e}$ or $N_\mathrm{e}$ \\
temden & getCrossTemDen & $T_\mathrm{e}$ and $N_\mathrm{e}$ \\
ntcontour & plotLineRatio, & Diagnostic plot \\
& plotContour & \\
redcorr & correctData & Dereddened fluxes \\
ionic  & getIonAbundance & Ionic abundances\\
\hline 
   \end{tabular}
      \end{table*}

The next section compares the output of selected NEBULAR tasks with the corresponding PyNeb's methods for specific cases. The NEBULAR version used to conduct the tests is the one contained in IRAF 2.16.

\subsection{Populations, emissivities, and physical conditions}

\begin{figure*}
   \centering
  \includegraphics[width=12cm]{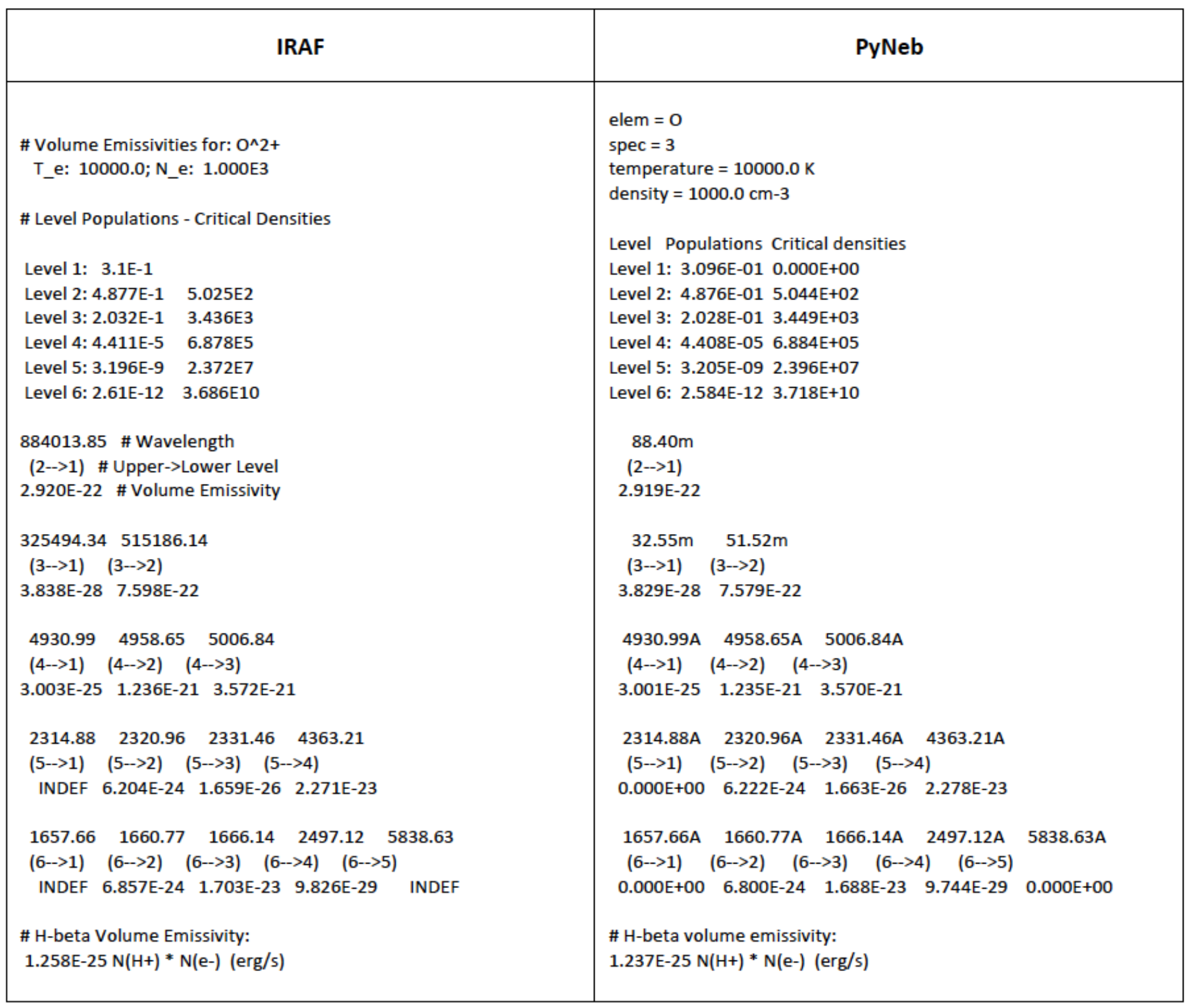}
   \caption{Comparison between the output of the NEBULAR task \texttt{ionic} and the PyNeb method \texttt{printIonic}, both run with the [O{\sc~iii}] data set adopted by NEBULAR in 2009.}
              \label{fig:printIonic}%
\end{figure*}

\begin{figure*}
   \centering
  \includegraphics[width=12cm]{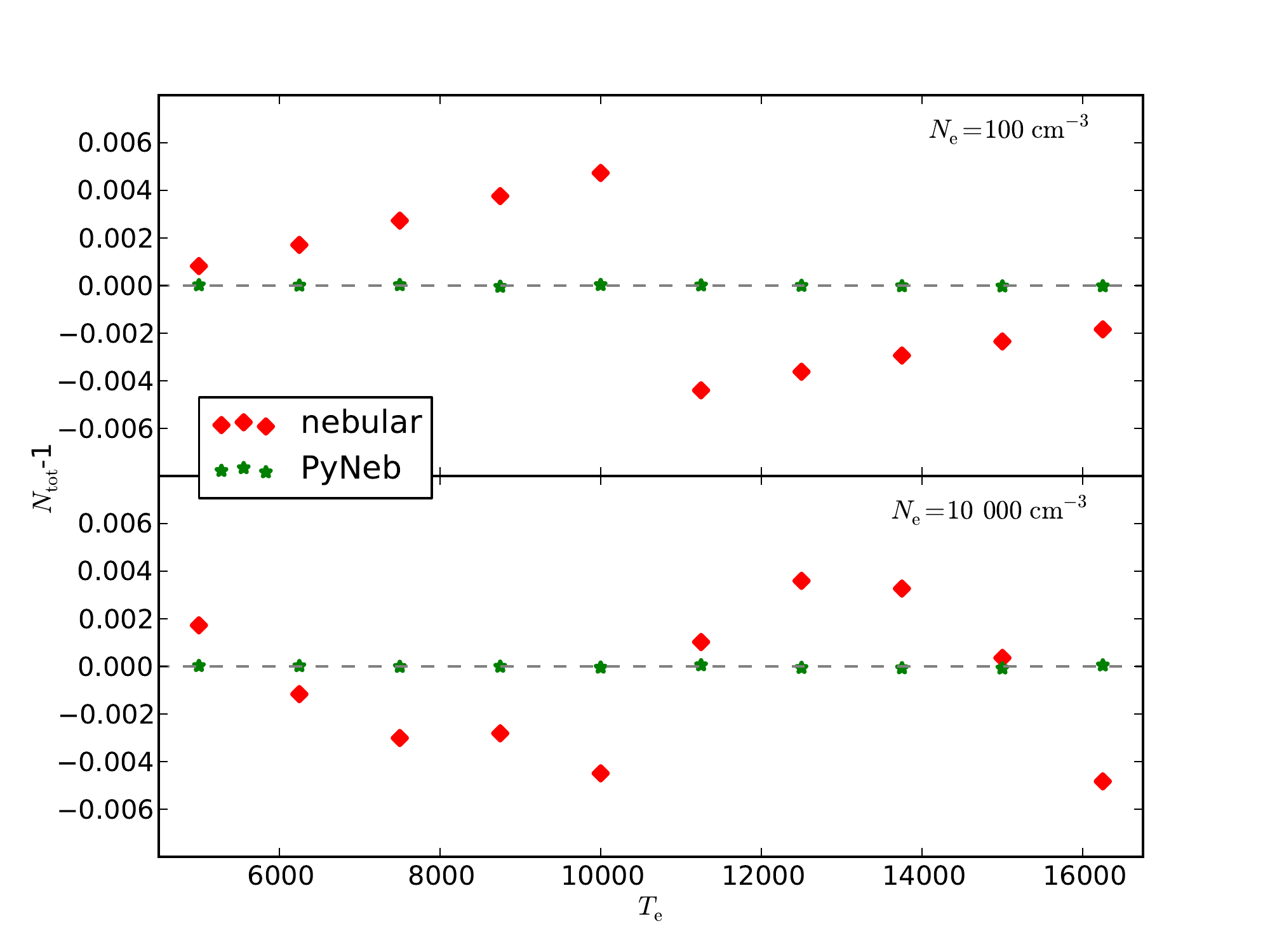}
   \caption{Normalization error in the total [S~{\sc ii}] population output by NEBULAR and Pyneb at two representative density values. }
              \label{fig:populations}%
\end{figure*}

Figure~\ref{fig:printIonic} compares the output of \texttt{ionic} and \texttt{printIonic}, both of which run with the same [O~III] data set. The output includes the level populations, the critical densities, and the emissivities. The results from the two codes agree in general but also exhibit slight differences. Even if these are negligible from a practical point of view, it is important to understand them to ensure that no systematic errors affect the output of either code. It should be noticed that all differences can ultimately be ascribed to differences between the computed level populations. While we could not track back the differences to any specific factor, the experiments conducted suggest that the differences are not systematic and are most probably related to the greater precision of both constants and variables enforced in PyNeb. This implies that the differences are negligible and can be safely ignored; furthermore, on the scale of the tiny differences found, PyNeb is more precise than NEBULAR. As an example, we show in Fig.~\ref{fig:populations} 
the error in the total [S{\sc~ii}] population output by both NEBULAR and PyNeb. 

\begin{figure*}
   \centering
  \includegraphics[width=12cm]{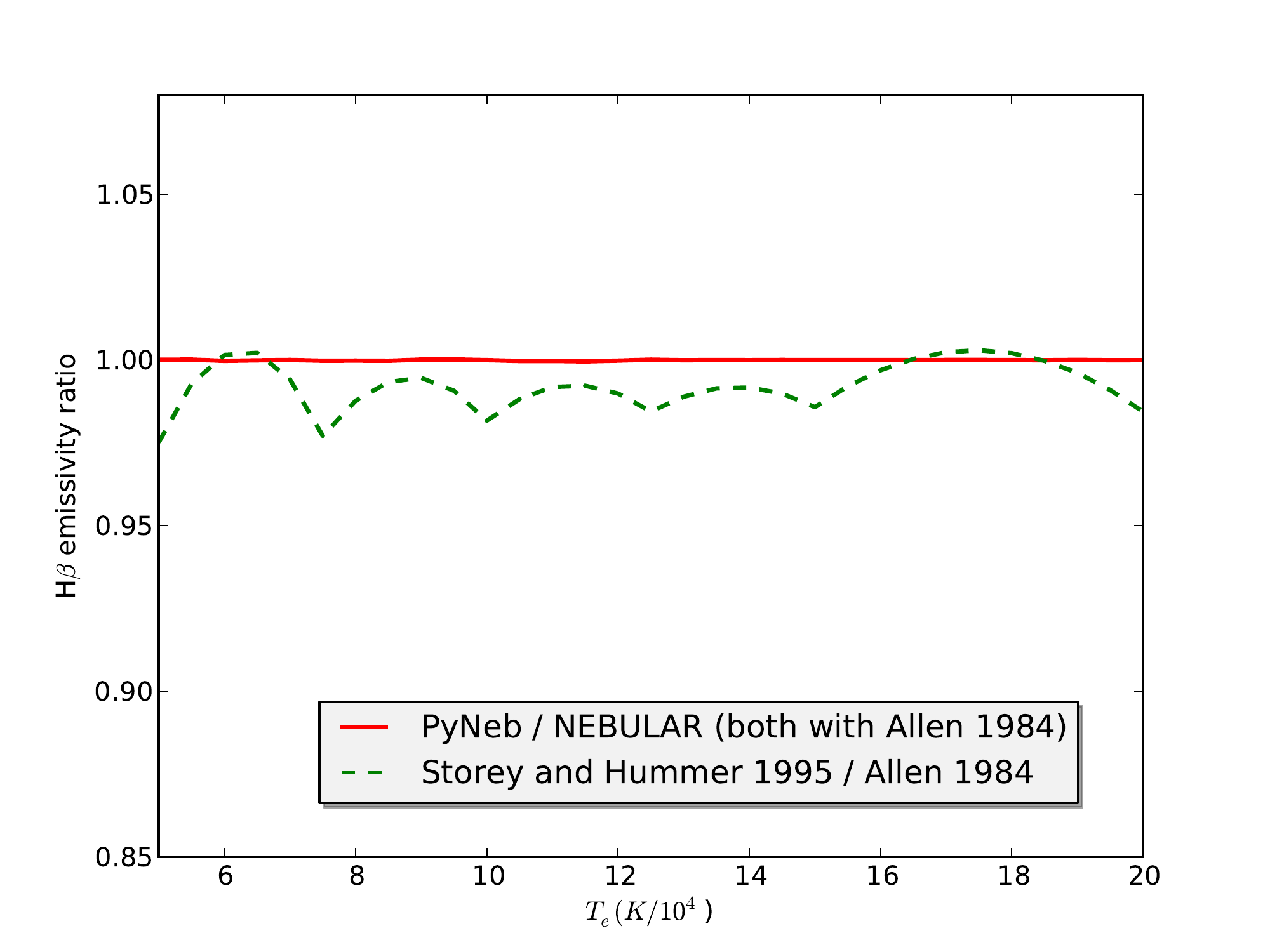}
   \caption{Comparison between the H$\beta$ emissivities computed by NEBULAR and PyNeb with the \cite{1984ASSL..112.....A} formula. Also shown is the ratio of PyNeb's default H$\beta$ emissivity, computed by linear interpolating in \cite{1995MNRAS.272...41S} tables, relative to the emissivity by cite{1984ASSL..112.....A}.}
              \label{fig:Hb_emissivity}%
\end{figure*}

Figure~\ref{fig:Hb_emissivity} compares the emissivity of H$\beta$ returned by NEBULAR with the one computed by PyNeb with the same interpolation formula \citep{1984ASSL..112.....A}.
The default emissivity used by PyNeb \citep{1995MNRAS.272...41S} is also displayed for reference, normalized to the \cite{1984ASSL..112.....A} formula. The wiggles are an artifact of the linear interpolation in the table; they indicate that a smooth interpolation would probably work better. Focusing on the data points underlying the wiggles, it can be seen that the H$\beta$ emissivity by \cite{1995MNRAS.272...41S} is roughly 2\% less than the one predicted by \cite{1984ASSL..112.....A}.

\begin{figure*}
   \centering
  \includegraphics[width=12cm]{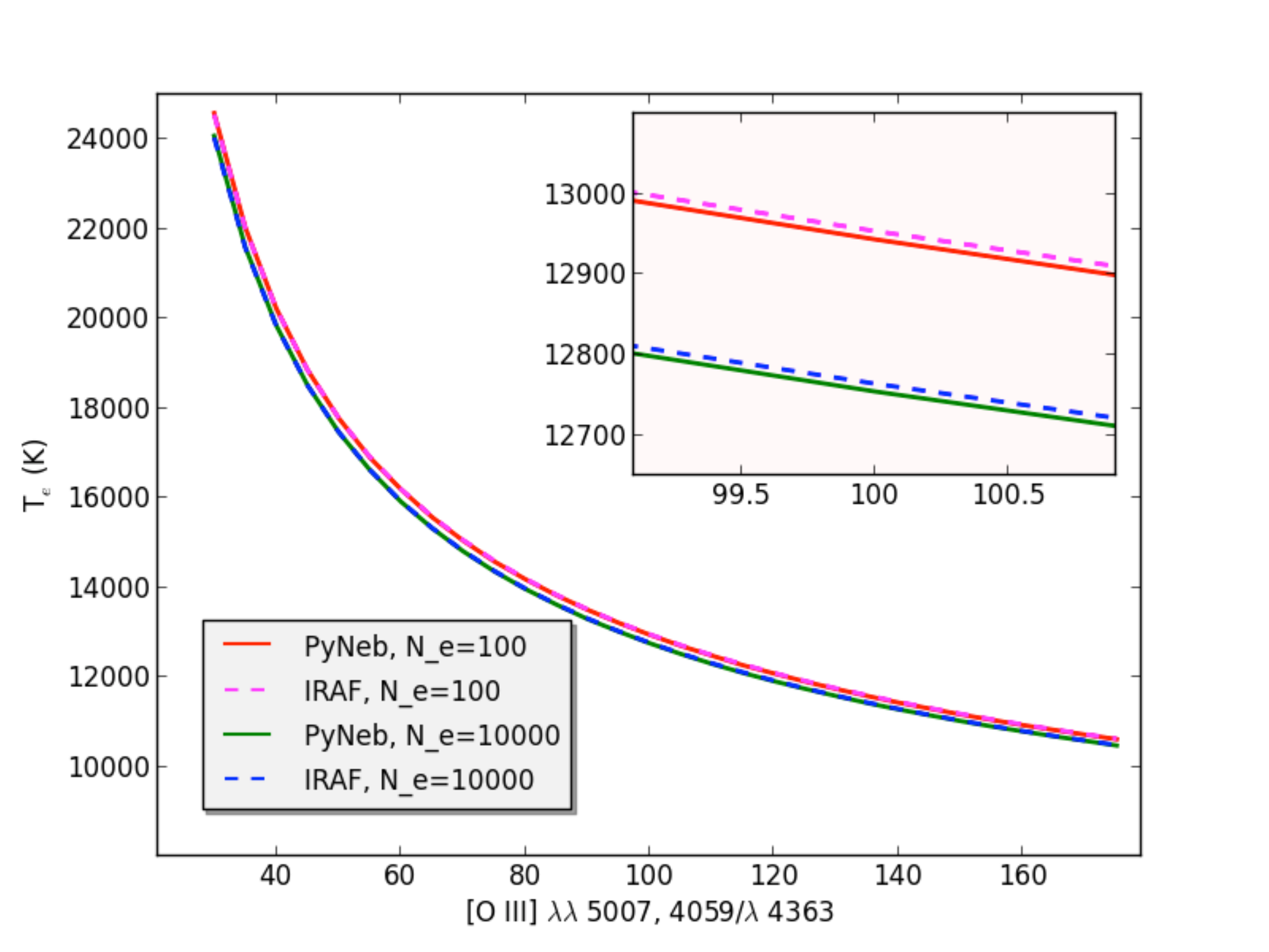}
   \caption{Comparison between the output of the NEBULAR task \texttt{temden} and the PyNeb method \texttt{getTemDen}, both run with the [O{\sc~iii}] data set adopted by IRAF in 2009. }
              \label{fig:temden}%
\end{figure*}

Figure~\ref{fig:temden} compares the [O{\sc~iii}]$\lambda\lambda 5007, 4959$/$\lambda 4363$ ratio
emissivity of H$\beta$ returned by NEBULAR's task \texttt{temden} with the one computed by PyNeb's method \texttt{getTemDen}, for two different density values. On the scale of the figure, NEBULAR's results are indistinguishable from PyNeb's results. By zooming in, differences on the order of 10 K can be seen. These differences, which are negligible in all practical applications, can be ascribed to the differences in the level populations mentioned above.
 
Tables~\ref{tab:diags_den} and \ref{tab:diags_tem} compare the results of NEBULAR and PyNeb for all the density and temperature diagnostics of Tables 2 and 3 of \cite{1995PASP..107..896S} at selected temperature and density points. Again, the differences are negligible in all cases.
 
   \begin{table*}
\centering 
      \caption[]{Comparison between density diagnostics in NEBULAR and PyNeb. Densities have been rounded off to the tens for values below $10^5$ cm$^{-3}$ and to the hundreds elsewhere.} 
      \label{tab:diags_den} 
\begin{tabular}{lllllrr}
\hline 
\hline 
Ion & Spectrum & Line Ratio & Value & Assumed $T_{\mathrm e}$ & $N_{\mathrm e}$ (NEBULAR) & $N_{\mathrm e}$ (PyNeb) \\
\hline 
N$^0$   & [N{\sc~i}]   & $I$(5198)/$I$(5200) & 2.00 & 10000 K &   3420 &   3420  \\
S$^+$   & [S{\sc~ii}]  & $I$(6716)/$I$(6731) & 0.75 & 10000 K &   1740 &   1740  \\
C$^+$   &  C{\sc~ii}]  & $I$(2326)/$I$(2328) & 1.60 & 10000 K &    100 &    100  \\  
O$^+$   & [O{\sc~ii}]  & $I$(3726)/$I$(3729) & 1.50 & 10000 K &   1100 &   1090  \\
Si$^{+2}$ & Si{\sc~iii}] & $I$(1883)/$I$(1892) & 0.10 & 10000 K & 772100 & 771300  \\ 
Cl$^{+2}$ & [Cl{\sc~iii}]& $I$(5517)/$I$(5538) & 0.75 & 10000 K &   6480 &   6470  \\  
N$^{+2}$  &  N{\sc~iii}] & $I$(1749)/$I$(1752) & 1.20 & 10000 K &    510 &    510  \\ 
Ar$^{+3}$ & [Ar{\sc~iv}] & $I$(4711)/$I$(4740) & 0.50 & 10000 K &  23250 &  23240  \\    
C$^{+2}$  & C{\sc~iii}]  & $I$(1907)/$I$(1909) & 0.40 & 15000 K & 115800 & 114400  \\    
O$^{+3}$  & O{\sc~iv}]   & $I$(1401)/$I$(1405) & 1.25 & 10000 K &   4640 &   4650  \\
Ne$^{+3}$ & [Ne{\sc~iv}] & $I$(2423)/$I$(2425) & 2.00 & 10000 K &  19400 &  19400  \\ 
\hline 
   \end{tabular}
      \end{table*}

   \begin{table*}
\centering 
      \caption[]{Comparison between temperature diagnostics in NEBULAR and PyNeb. Temperatures have been rounded off to the tens.}  
      \label{tab:diags_tem} 
\begin{tabular}{lllllrr}
\hline 
\hline 
Ion & Spectrum & Line ratio & Value & Assumed $N{\mathrm e}$ & $T_{\mathrm e}$ (NEBULAR) & $T_{\mathrm e}$ (PyNeb) \\
\hline 
O$0$   & [O{\sc~i}]    & $I$(5577)/$I$(6300+6363)      & 0.01  & 1000  &    8400 &  8400  \\
S$+$   & [S{\sc~ii}]   & $I$(4068+4076)/$I$(6716+6731) & 0.10  & 100   &   11510 & 11510  \\
O$+$   & [O{\sc~ii}]   & $I$(3726+3729)/$I$(7320+7330) & 100   & 100   &    7480 &  7480  \\
N$+$   & [N{\sc~ii}]   & $I$(5755)/$I$(6548+6583)      & 0.1   & 100   &    9970 &  9970  \\
Si$+2$ & Si{\sc~iii}]  & $I$(1206)/$I$(1883+1892)      & 0.025 & 1000  &   15790 & 15790  \\ 
S$+2$  & [S{\sc~iii}]  & $I$(6312)/$I$(9069+9532)      & 0.02  & 1000  &  11040  & 11090  \\
Ar$+2$ & [Ar{\sc~iii}] & $I$(5192)/$I$(7136+7751)      & 0.01  & 1000  &   12410 & 12420  \\    
O$+2$  & [O{\sc~iii}]  & $I$(4363)/$I$(4959+5007)      & 0.01  & 1000  &   12940 & 12920  \\
Cl$+3$ & [Cl{\sc~iv}]  & $I$(5323)/$I$(7530+8045)      & 0.01  & 1000  &    8180 &  8180  \\
Ar$+3$ & [Ar{\sc~iv}]  & $I$(2854+2868)/$I$(4711+4740) & 0.01  & 100   &    8270 &  8270  \\   
Ne$+2$ & [Ne{\sc~iii}] & $I$(3342)/$I$(3869+3969)      & 0.001 & 1000  &   10050 & 10060  \\
Ar$+4$ & [Ar{\sc~v}]   & $I$(4626)/$I$(6435+7006)      & 0.01  & 1000  &   12520 & 12500  \\   
Ne$+4$ & [Ne{\sc~v}]   & $I$(2975)/$I$(3426+3346)      & 0.001 & 1000  &   12020 & 12040  \\
\hline 
   \end{tabular}
      \end{table*}

\begin{figure*}
   \centering
      \subfigure[]{%
            \label{fig:ntcontour_first}
           \includegraphics[width=0.4\textwidth]{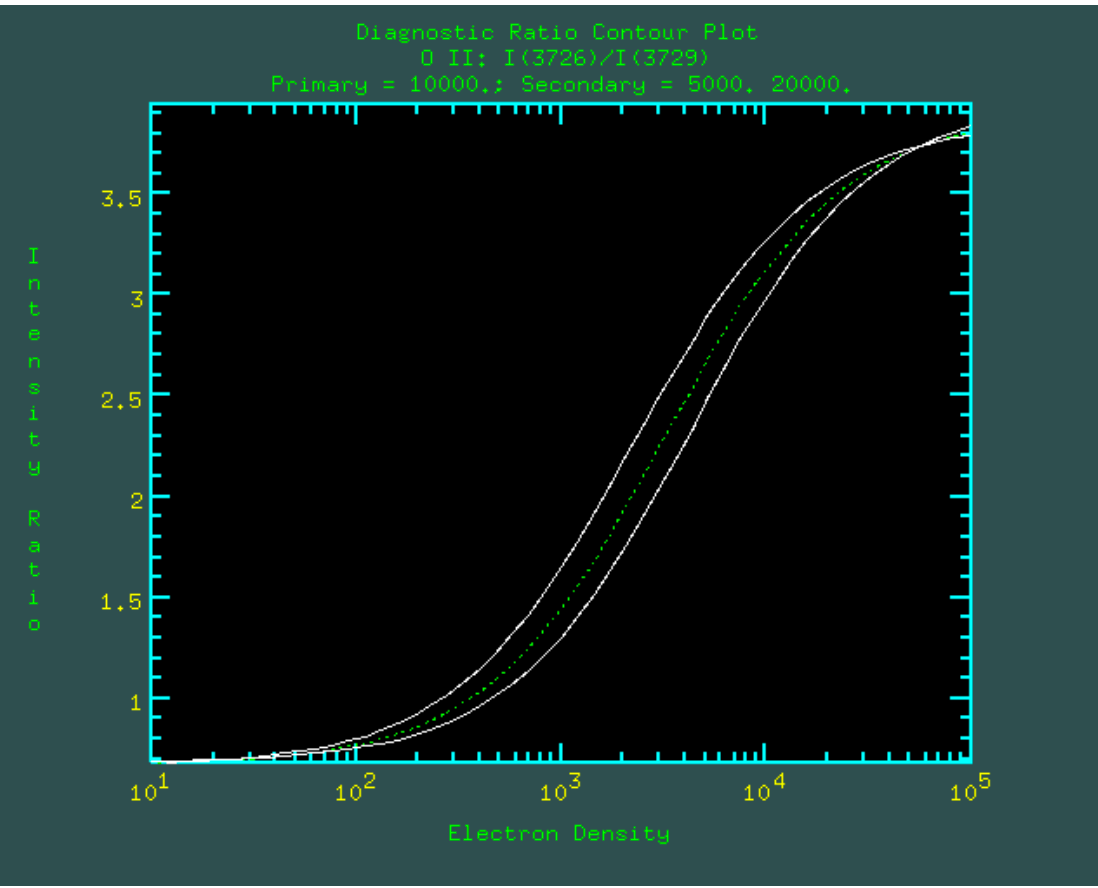}
        }%
      \subfigure[]{%
            \label{fig:ntcontour_second}
              \includegraphics[width=0.475\textwidth]{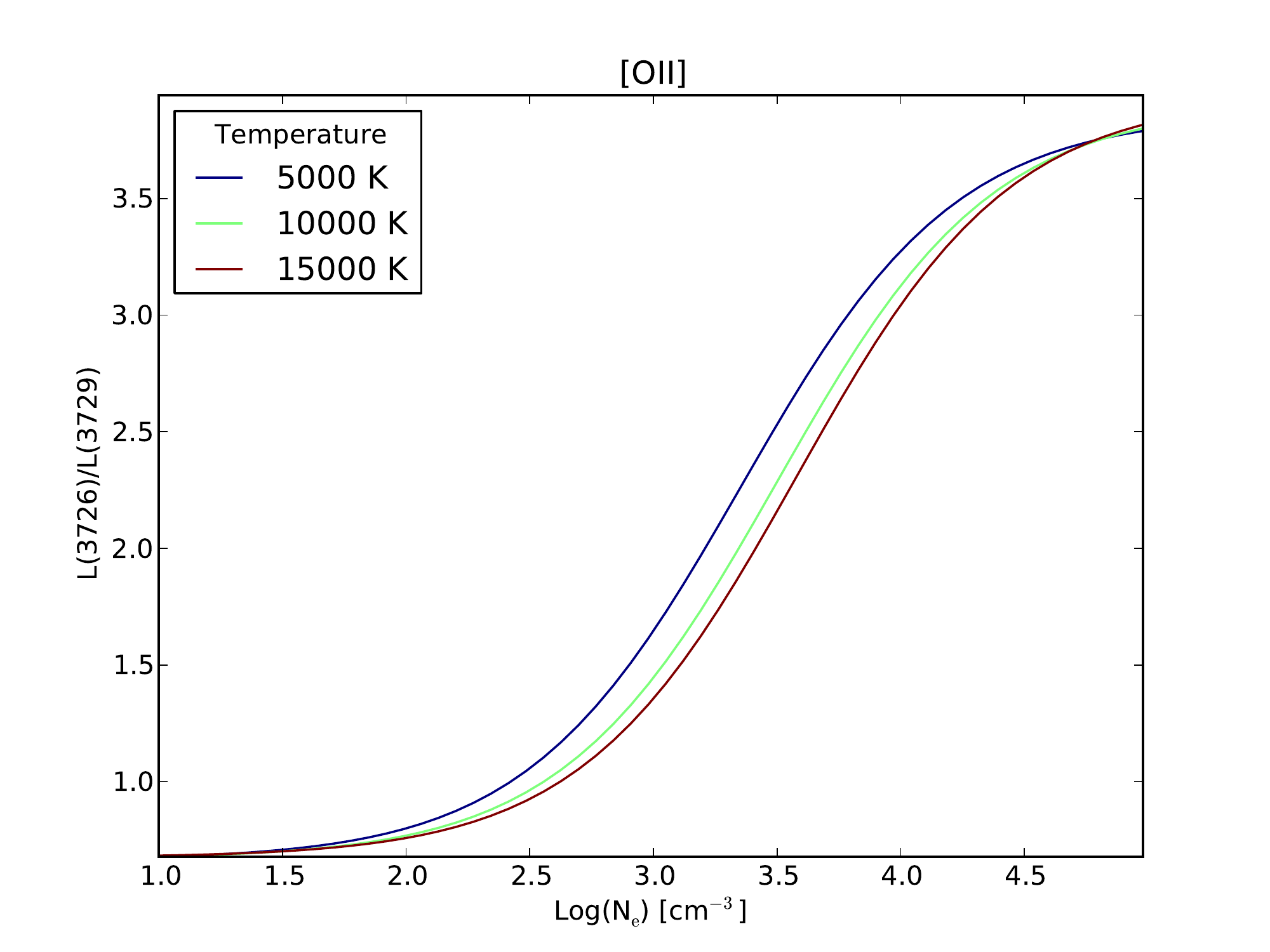}
              }
   \caption{Left: Density diagnostic plot of the [O{\sc~iii}] $\lambda$3726/$\lambda$3729 intensity ratio as a function of electron density and temperature, obtained with the NEBULAR command \texttt{ntcontour}. Right: density diagnostic plot of the [O{\sc~iii}] $\lambda$3726/$\lambda$3729 intensity ratio as a function of electron density and temperature, obtained with the PyNeb command \texttt{plotLineRatio}.}
              \label{fig:ntcontour}%
   \end{figure*}

Finally, diagnostic plots can be used in essentially the same way in both NEBULAR and PyNeb, with the latter allowing a greater flexibility in terms of the plot's graphical features (Fig.~\ref{fig:ntcontour}).
 
\subsection{Limitations and caveats}\label{sec:limitations}

Powerful as it is, PyNeb incorporates a fairly simple physical model of line emission in nebular objects, and it is important to ensure that it is not stretched outside its range of validity. This section enumerates a few physical problems that cannot be solved by PyNeb.

An important process that cannot be accounted for by PyNeb is fluorescence, which may affect the emitted intensity of both collisional and recombination lines \citep{2005MNRAS.361..813E, 2009ApJ...691.1712L}.  Fluorescence is a line emission mechanism triggered by the absorption of a photon previously emitted either from a stellar source (continuum pumping) or from another ion (resonant fluorescence). 
While the intensity of collisional and recombination lines merely depends on the local conditions and the abundance of the ion involved, the intensity emitted through fluorescence depends on the intensity, spectrum, position, and relative velocity of the stellar source (in the case of continuum pumping) or on the simultaneous emission of other ions (fluorescence of resonant lines).
The intensity of the line produced through this process adds to the intensity produced by collisional or recombination processes, a fact to take into account when observational data are analyzed. Since PyNeb considers the ions in isolation, it cannot account for this line-enhancing process, which must be corrected for explicitly before interpreting observed intensities in terms of physical conditions and chemical abundances.

In the context of fluorescence, it is useful to discuss the recent results by \citet{2014A&A...561A..10P}, who claim that the usual relation between the electron temperature and the [O~{\sc iii}] line ratio $R$=$\lambda\lambda$(5007+4959)/$\lambda$4363  (as computed, for instance, by PyNeb) is biased and should be revised. 
The new relation they propose would lead to a decrease in the electron temperature computed from a given ratio $R$, compared to what would be obtained with the standard relation, with the discrepancy between both relations increasing with the electron temperature (see their Fig.1). For several reasons, we consider these results to be erroneous. First of all, the authors claim that the effect is due to fluorescence, but they fail to notice that, in their Cloudy photoionization models \citep{1998PASP..110..761F}, none of the three lines involved in the ratio $R$ is affected by this mechanism. Second, since their results are based on a solar-metallicity grid, the authors are forced to increase the effective temperature of the ionizing SED up to unrealistic values of a few 10$^5$ K in order to reach high electron temperatures; furthermore, the models are density-bounded and have a very small geometrical thickness. As a consequence, the models with the highest electron temperature also have a very high ionization degree, which leads to unrealistically high values of O$^{3+}/$O$^{2+}$ of  $\sim$100. This, in turn, causes a huge increase in the contribution of O$^{3+}$ recombination to the emissivity of the [O{\sc~iii}] $\lambda$4363 line (see Eq. 3 of \cite{2000MNRAS.312..585L}) by the same factor and, consequently, an artificial decrease in the diagnostic line ratio at a given electron temperature. In real nebulae, the contribution of the recombination to this line is invariably much smaller than the contribution determined by \citet{2014A&A...561A..10P} (classical planetary nebulae only marginally reach 1 and H{\sc~ii} regions have even lower values), ruling out the validity of their Eq.~1. If one ever needs to take into account the contribution of recombination to the intensity of the  [O~{\sc iii}] $\lambda$4363 line, we suggest using the recipe by \citet{2000MNRAS.312..585L}.

Similarly, PyNeb is a {\it local} code, so it is not designed to solve radiation transfer. 
This does not hinder interpretation of observed lines since, when optical depths are low --which is true for most of these lines-- photons freely escape, and there is no need of a detailed treatment of radiation transfer.  Dust is only included as the source of foreground extinction, but 
PyNeb is not designed to compute the effect of dust grains on the thermal or ionization equilibrium.

PyNeb cannot account for energy balance issues either (thus, it cannot give clues to the temperature fluctuations problem), nor can it compute a finite-depth, complex physical model. 
In short, PyNeb is not a photoionization code but rather a code for analyzing emission lines.

Finally, it must be remembered that both the $A$s and the $\Upsilon$s are subject to uncertainties. The discussion of these and the description of PyNeb's atomic data base will be the subject of Paper~II.

\section{Practical details}\label{sec:using}

\subsection{Software requirements}

As of mid 2014,
the code requires the following packages and libraries to function:

\begin{enumerate}
\item{} {\it python}\footnote{http://www.python.org} v. 2.7x ({\it{not}} v. 3.x, which is a different and incompatible branch of python)
\item{} {\it numpy}\footnote{http://www.numpy.org}  v. 1.6 or later 
\item{} {\it matplotlib}\footnote{http://www.matplotlib.org} v. 1.2 or later
\item{} {\it scipy}\footnote{http://www.scipy.org} v. 0.10 or later.
\end{enumerate}

The first two are required for the code to function. The graphical library \textit{matplotlib} is needed to use the graphic modules of PyNeb, but the code is prepared
to skip them if the library is not installed. In such cases, a warning will be issued, but the code will not stop.
Finally, {\it scipy} is currently only required to compute recombination emissivities with the \texttt{RecAtom} class (see Sect.~\ref{sec:RecAtom}) and to compute stellar temperatures with the \texttt{Stellar} class.

\subsubsection{Running the code}

PyNeb can be used in two ways: either interactively feeding individual commands to the command line or running a script. The first way to proceed is suitable for quick, simple computations; the second is more adequate if a large number of commands have to be executed in a sequence. Examples of both are given in the documentation, which is discussed in Sect.~\ref{sec:documentation}.

\subsubsection{Running with multiple processors}

PyNeb takes full advantage of {\it numpy}'s features to perform fast, vectorized computations, thus avoiding time-consuming loops in repetitive operations such as computing an emissivity grid, which requires computing an $n_\mathrm{tem} \times n_\mathrm{den}$ array of emissivities. In addition to this, the method {\bf Atom.getTemDen} (which computes the temperature given the density and vice versa, and will be presented in Sect.~\ref{sec:Atom}) has been parallelized to be used with multiple processors, reducing the execution time further. The method \texttt{Diagnostics.getCrossTemDen}, which simultaneously computes the temperature and density and will be described in Sect.\ref{sec:diagnostics}, works by iteratively calling  \texttt{Atom.getTemDen} and is therefore also boosted by the use of multiple processors.

\subsection{Documentation and support}\label{sec:documentation}

Efficient and safe use of a public package is only possible if the package is well documented. PyNeb currently has several sources of documentation, also listed on the web page of the code at \url{http://www.iac.es/proyecto/PyNeb}:

\begin{enumerate}
\item{} PyNeb's manual: a short, discursive introduction to PyNeb with plenty of examples, which is continuously updated as the code evolves.  
\item{} The code docstrings, which can be read interactively through standard python or ipython commands: these are also used to automatically generate a reference document, which is available in both html and pdf formats.
\item{} A series of well-commented scripts, which illustrate real science cases and can be downloaded from the web page.
\item{} This paper and Paper II.
\end{enumerate}

In addition, there is also a Google discussion group called PyNeb\footnote{\url{https://groups.google.com/forum/\#!forum/pyneb}}  to post requests, help other users, and share your problems. Direct support from the development team (\email{pyneb@googlegroups.com}) is on a time-available basis.

\section{Concluding remarks: The present and future of PyNeb}\label{sec:conclusions}

The following is a list that summarizes  PyNeb's main features:

\begin{enumerate}
\item{} A large sample of atomic data is provided with the code. More are being added, and tools to add your own data are also provided. Changing from one atomic data set to another is foolproof. Pyneb also includes tools to explore and visualize the atomic data used. 
\item{} Thanks to {\it matplotlib}, PyNeb has improved graphical capabilities to produce print quality plots. Additionally, because the user has full access to intermediate quantities of computations, built-in commands can be complemented by user-produced scripts to produce tailored plots.
\item{} Simultaneous determination of temperature and density from pairs of line ratios is possible without the need of external looping on \texttt{getTemDen}.
\item{} Elemental abundances can be computed from ionic abundances with a simple command. A selection of published ICFs is provided with complete information on their analytical form and bibliographic source. The collection of built-in ICFs can be complemented by user-provided formulae. 
\item{} Recombination intensities for some important ions have been added.
\item{} Some ions have been added to the NEBULAR inventory of collisional atoms, including both common ions (such as [Cl{\sc~ii}]) and several s-elements.
\item{} Insight into emissivity properties is possible thanks to emissivity grids.
\item{} PyNeb has a modular structure, which makes it easy to embed modules in scripts.
\end{enumerate}

Although the core functionality of PyNeb is already in place and fairly stable, we continue to improve the code and to add new functionality. The following are some areas of future potential development:

\begin{enumerate}
\item{} We intend to add new ions to the inventory of collisional-excited lines; in particular, we are seeking to extend the inventory of $s$-process elements and other heavy ions.  More details on this point are given in Paper~II.
\item{} We also plan to extend the choice of recombination ions to include, e.g., O{\sc~ii} and C{\sc~ii}.
\item{} Currently, the treatment of blends is not full: while they can be included in diagnostics, they cannot be used yet as input to the \texttt{getIonAbundance} method. We intend to improve this in the near future.
\item{} A Monte Carlo procedure to estimate the error in abundances derived from errors in fluxes will be added to the code.
\end{enumerate}

Ideally, it would also be desirable to be able to create VO-compatible web services to provide PyNeb over the internet, or, alternatively, to create a GUI to be installed locally to provide at least some of PyNeb's functionality. While we currently lack the labor-force to accomplish such goals, we do not exclude that they might be achieved in the future, possibly with the collaboration of users.   We would be grateful for any feedback, query, or comment on the code and its accompanying documentation. 

PyNeb has already been used in a number of papers \citep{2012A&A...538A..54G, 2013A&A...558A.122G, 2014arXiv1410.0119G, 2013MNRAS.433..382E, 2014MNRAS.443..624E, 2013MNRAS.431..159K, 2013ApJS..207...21N, 2014ApJ...786..155N, 2013A&A...559A.114M, 2014MNRAS.440..536D, 2014ApJ...784..173D, 2014ApJ...785..100M, 2014ApJ...787....3S, 2014arXiv1407.2411H, 2014MNRAS.441.2663P}.
The scientific community is invited to keep using it for their projects. Whenever you use PyNeb for calculations that lead to a published paper, you are kindly asked to cite both this paper and Paper~II.

\begin{acknowledgements}
 VL acknowledges support of the Spanish Ministry of Science and Innovation through grant AYA 2011-22614.
 CM acknowledges support from CONACyT project CB-2010/153985.
Support for programs numbered HST-GO--11657 and HST-GO--112600 was provided by NASA through a grant from the Space Telescope Science Institute, which is operated by the Associated Universities for Research in Astronomy, Inc., under NASA contract NAS5-26555. 
Thanks are due to Amanda Mashburn and Nick Sterling for kindly providing data for several $s$-process ions. We also acknowledge useful suggestions from Jorge Garc{\'\i}a-Rojas, Adal Mesa Delgado, Luis L\'opez-Mart{\'\i}n and Rub\'en Garc{\'\i}a-Benito.
 \end{acknowledgements}



\bibliographystyle{aa}
%
\bibliography{pyneb_refs}{}

\begin{appendix}
\section{Sample scripts}\label{sec:sample}

\begin{figure*}
   \centering
   \includegraphics[width=12cm]{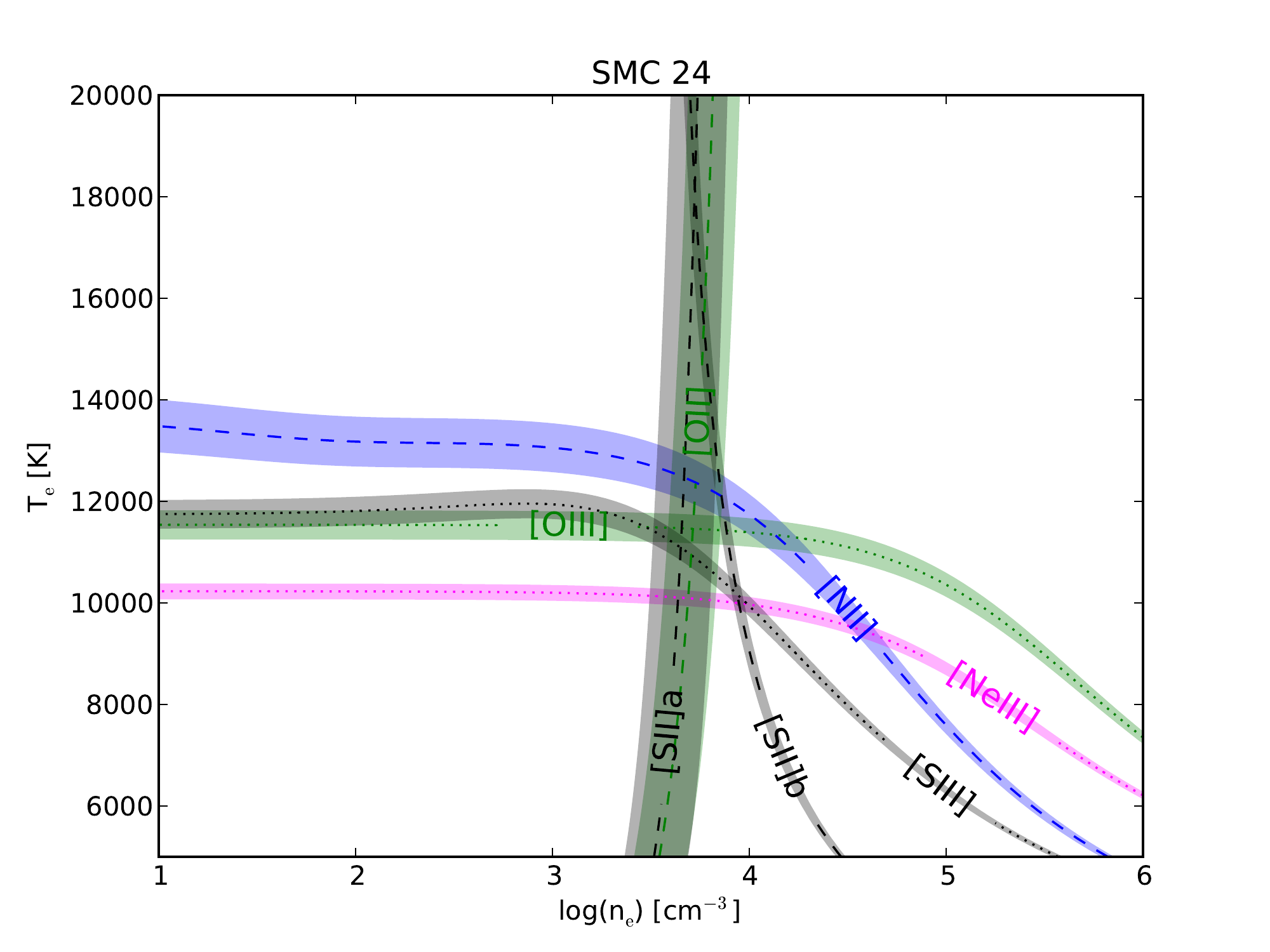}
   \caption{Diagnostic diagram of SMC24 produced with smc24\_diagnostics.py.}
              \label{fig:smc24}
   \end{figure*}

Here
we include two sample scripts that illustrate how real physical problems can be analyzed with PyNeb. The first script produces a diagnostic plot similar to the one of Fig.~\ref{fig:ic418} (Fig.~\ref{fig:smc24}). The content of the data file is given at the end of the script and can also be downloaded from http://www.iac.es/proyecto/PyNeb/smc24.dat. 
\small
\begin{verbatim}

# PyNeb script smc24_diagnostics.py

# Imports
import pyneb as pn
import matplotlib.pyplot as plt

# Adopt an extinction law
extinction_law = 'CCM 89'

# Define the data file 
obs_data = 'smc24.dat'

# Plot title
title = 'SMC 24'

### Read and deredden observational data
# define an Observation object and name it 'obs'
obs = pn.Observation()

# fill obs with data from obs_data
obs.readData(obs_data, fileFormat='lines_in_rows', 
                     err_default=0.05)

# deredden data with Cardelli's law
obs.extinction.law = extinction_law
obs.correctData()

### Include the diagnostics of interest
# instantiate the Diagnostics class
diags = pn.Diagnostics()
# include in diags the relevant line ratios
diags.addDiag([
              '[NII] 5755/6584', 
              '[OII] 3726/3729', 
              '[OIII] 4363/5007', 
              '[SII] 6731/6716', 
              '[SII] 4072+/6720+',
              '[SIII] 6312/18.7m', 
              '[NeIII] 3930+/15.6m', 
              ])
diags.addClabel('[SII] 6731/6716', '[SII]a')
diags.addClabel('[SII] 4072+/6720+', '[SII]b')

# Create the emission maps to be compared to
# the observation data (some overkill here)
emisgrids = pn.getEmisGridDict(atom_list=\
    diags.getUniqueAtoms(), den_max=1e6)

### Plot
# Create the contour plot as the intersection of 
# tem-den emission maps with dereddened line ratios
diags.plot(emisgrids, obs)

# Place the title
plt.title(title)

# Display the plot
plt.show()

# Data
# NAME SMC_24    
# cHbeta       0.047  
# S4_10.5m     7.00000  
# Ne2_12.8m    8.3000  
# Ne3_15.6m   34.10  
# S3_18.7m    10.4  
# O2_3726A    39.700  
# O2_3729A    18.600  
# Ne3_3869A   18.90  
# Ne3_3967A    6.4  
# S2_4069A     0.85  
# S2_4076A     0.450  
# O3_4363A     4.36  
# O3_5007A     435.09 
# N2_5755A     0.510000  
# S3_6312A     0.76  
# O1_6300A     1.69  
# O1_6364A     0.54  
# N2_6548A     6.840000  
# N2_6584A    19.00  
# S2_6716A     1.220000  
# S2_6731A     2.180000  
# Ar3_7136A    4.91  
# O2_7319A+    6.540000  
# O2_7330A+    5.17  

\end{verbatim}
\normalsize

\begin{figure*}
   \centering
   \includegraphics[width=12cm]{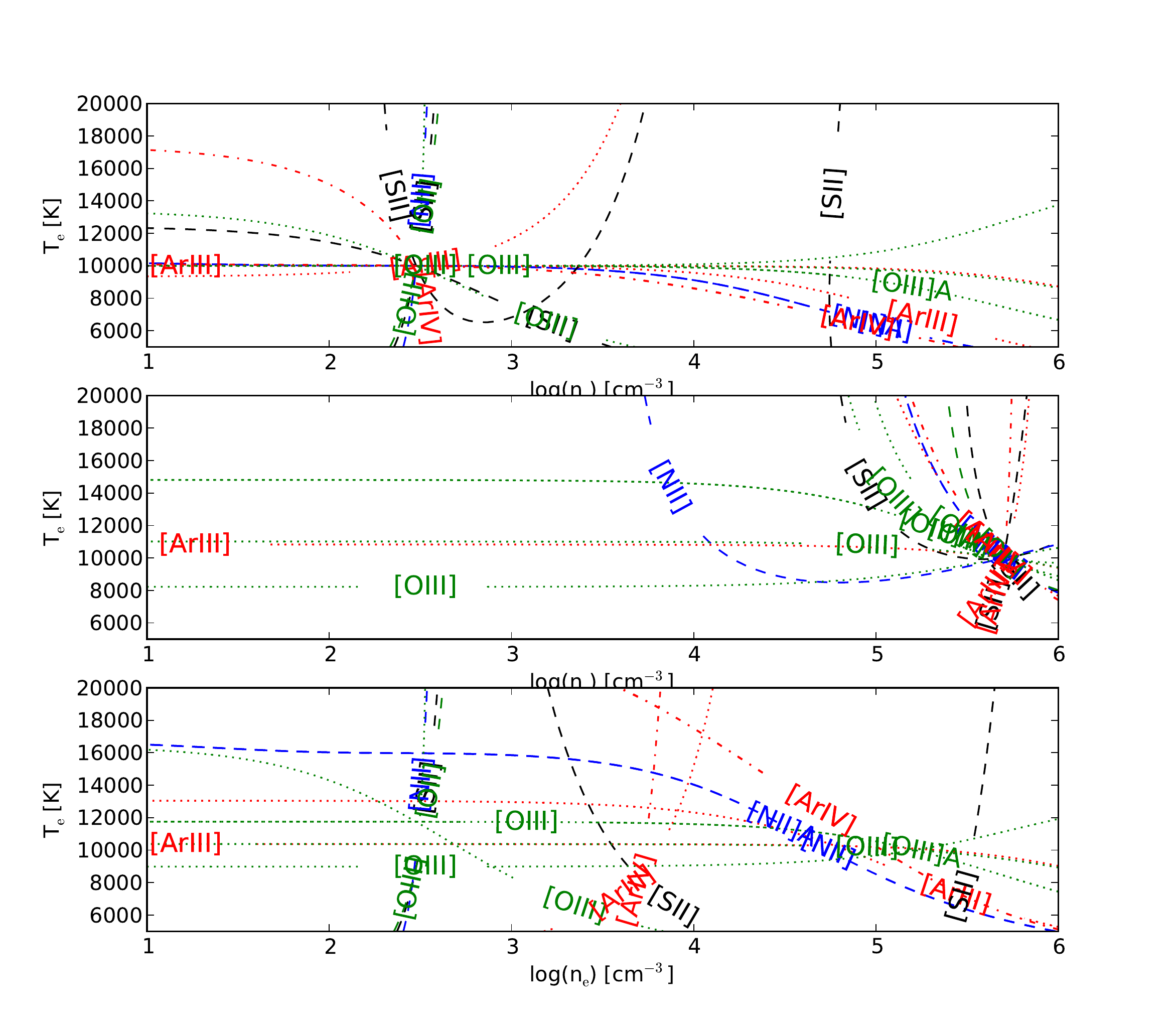}
   \caption{Plots produced with the script two\_comp.py to illustrate the bias resulting from analyzing a two-density region as a homogeneous one.}
              \label{fig:two_comp}%
   \end{figure*}

The second script illustrates the case of a density-inhomogenous region that is erroneously analyzed as a constant-density region (see also Fig.~\ref{fig:two_comp}).

\small
\begin{verbatim}

# PyNeb script smc24_diagnostics.py

# imports
import pyneb as pn
import matplotlib.pyplot as plt

# Define the two subregions in terms 
# of temperature, density and mass

tem1 = 1e4
tem2 = 1e4
dens1 = 3e2
dens2 = 5e5
mass1 = 1
mass2 = 5e-4

# Create list of diagnostics to analyze the region
diags = pn.Diagnostics()
for ion in ['O2', 'O3', 'S2', 'N2', 'Ar3', 'Ar4']:
    diags.addDiag(atom=ion)
    print 'Adding diagnostics of', ion

# Use alternate labels for some ratios
diags.addClabel('[NII] 5755/6584', '[NII]A')
diags.addClabel('[OIII] 4363/5007', '[OIII]A')

# Create all the ions involved in the diagnostics
adict = diags.atomDict
pn.log_.message('Atoms built')

# Create a virtual observation with 3 records:
# subregion1, subregion 2 and the sum of both
obs = pn.Observation(corrected = True)

# Compute the intensities of all the lines
# of all the ions considered
for atom in adict:
    for line in pn.LINE_LABEL_LIST[atom]:
        if line[-1] == 'm':
            wave = float(line[:-1])*1e4
        else:
            wave = float(line[:-1])
         
        intens1 = adict[atom].getEmissivity(tem1, \
                       dens1, wave=wave)*dens1*mass1
        intens2 = adict[atom].getEmissivity(tem2, \
                       dens2, wave=wave)*dens2*mass2
        elem = adict[atom].elem
        spec = adict[atom].spec
        obs.addLine(pn.EmissionLine(elem, spec, wave, \
              obsIntens= [intens1, intens2, \
              intens1+intens2], obsError=[0., 0., 0.]))
pn.log_.message('Virtual observations computed')

# Compute emission grids of the atoms considered
emisgrids = pn.getEmisGridDict(atomDict=adict)
pn.log_.message('EmisGrids available')

# Produce a diagnostic plot for each region and
# another one for the (misanalyzed) overall region
plt.figure()
plt.subplot(3, 1, 1)
diags.plot(emisgrids, obs, i_obs=0)
plt.subplot(3, 1, 2)
diags.plot(emisgrids, obs, i_obs=1)
plt.subplot(3, 1, 3)
diags.plot(emisgrids, obs, i_obs=2)
plt.show()

\end{verbatim}
\normalsize

%

\end{appendix}

\end{document}